\documentclass[fleqn,10pt]{wlscirep}
\usepackage[utf8]{inputenc}
\usepackage[T1]{fontenc}
\usepackage{subcaption} 
\usepackage{tabularx}

\title{REAL-Colon: A dataset for developing real-world AI applications in colonoscopy}

\author[1,*]{Carlo Biffi}
\author[2]{Giulio Antonelli}
\author[3]{Sebastian Bernhofer}
\author[4,5]{Cesare Hassan}
\author[6]{Daizen Hirata}
\author[6]{Mineo Iwatate}
\author[3]{Andreas Maieron}
\author[1]{Pietro Salvagnini}
\author[1,7,*]{Andrea Cherubini}

\affil[1]{Cosmo Intelligent Medical Devices, Dublin, Ireland}
\affil[2]{Gastroenterology and Digestive Endoscopy Unit, Ospedale dei Castelli (N.O.C.), Ariccia, Italy}
\affil[3]{Gastroenterology and Hepatology and Rheumatology, University Hospital of St. Pölten, St. Pölten, Austria}
\affil[4]{Department of Biomedical Sciences, Humanitas University, Pieve Emanuele, Italy}
\affil[5]{Endoscopy Unit, Humanitas Clinical and Research Center IRCCS, Rozzano, Italy}
\affil[6]{Gastrointestinal Center, Sano Hospital, Hyogo, Japan}
\affil[7]{Milan Center for Neuroscience, University of Milano–Bicocca, Milano, Italy}

\affil[*]{corresponding author(s): Andrea Cherubini (acherubini@cosmoimd.com), Carlo Biffi (cbiffi@cosmoimd.com)}

\begin{abstract}
Detection and diagnosis of colon polyps are key to preventing colorectal cancer. Recent evidence suggests that AI-based computer-aided detection (CADe) and computer-aided diagnosis (CADx) systems can enhance endoscopists' performance and boost colonoscopy effectiveness. However, most available public datasets primarily consist of still images or video clips, often at a down-sampled resolution, and do not accurately represent real-world colonoscopy procedures. We introduce the REAL-Colon (Real-world multi-center Endoscopy Annotated video Library) dataset: a compilation of 2.7M native video frames from sixty full-resolution, real-world colonoscopy recordings across multiple centers. The dataset contains 350k bounding-box annotations, each created under the supervision of expert gastroenterologists. Comprehensive patient clinical data, colonoscopy acquisition information, and polyp histopathological information are also included in each video. With its unprecedented size, quality, and heterogeneity, the REAL-Colon dataset is a unique resource for researchers and developers aiming to advance AI research in colonoscopy. Its openness and transparency facilitate rigorous and reproducible research, fostering the development and benchmarking of more accurate and reliable colonoscopy-related algorithms and models.
\end{abstract}

\begin{document}

\flushbottom
\maketitle

\thispagestyle{empty}

\section*{Background \& Summary}
Colorectal cancer (CRC) remains a significant global health concern, with approximately two million new cases detected annually\cite{sung2021global, morgan2023global}. Since more than 95\% of CRC originates from premalignant (adenomas) polyps, their detection and removal can substantially reduce the incidence and mortality of CRC\cite{bretthauer2022effect, zorzi2023adenoma}. Colonoscopy, a well-established screening procedure, has positively impacted CRC incidence in countries where it has been introduced\cite{dekker_advances_2018}. However, the inherent variability in the quality of colonoscopy due to its high dependence on human skill and vigilance poses challenges to its effectiveness as a screening tool\cite{kaminski_optimizing_2020, cherubini_gorilla_2023}.

In response to these challenges, Artificial Intelligence (AI) has emerged as a promising tool for augmenting human capabilities during live colonoscopy procedures\cite{ahmad2019artificial}. Its potential as a reliable tool for improving performances and standardizing screening is increasingly being recognized \cite{berzin_position_2020}. Evidence from randomized controlled trials highlights the effectiveness of computer-aided polyp detection (CADe) systems in preventing missed polyps, thereby enhancing the quality of colonoscopy procedures \cite{repici_efficacy_2020, wallace_impact_2022}. Furthermore, interest in computer-aided diagnosis (CADx) has grown, given its potential to assist real-time decision-making on polyp optical characterization \cite{spadaccini_computer-aided_2021, biffi_novel_2022}.

Significant challenges remain despite the promising advancements in this field, driven primarily by large MedTech corporations. The high cost and logistical complexities of acquiring and labeling large video recording datasets have limited the involvement of academic centers, with the scientific projects stemming from universities focusing primarily on small-size collections of still images or short video clips\cite{bernal2012towards,silva2014toward,bernal2015wm,tajbakhsh2015automated,angermann2017towards,mesejo2016computer,jha2020kvasir,sanchez2020piccolo,li2021colonoscopy,misawa2021development,ma2021ldpolypvideo,ali2023multi}. However, colonoscopy CAD(e/x) systems require the integration of video understanding algorithms that can execute a range of computer vision functions, including image and video classification, object classification, detection, and segmentation. Tabulated in Table\ref{tab:1} and Table\ref{tab:2} is a summary of available datasets for open research for polyp localization and classification. Only two datasets include information on polyp size,  no dataset is dedicated to polyp tracking and no dataset offers a comprehensive annotation of full, unaltered colonoscopy procedures. Instead, open datasets often focus on short video clips that include frames with polyps and omit extensive portions of colonoscopy videos without polyps (negative frames), or they sample only a small fraction of these negative frames. In contrast, the reality of colonoscopy videos features 80-90\% negative frames (as also illustrated in Figure \ref{fig:hist_boxes_per_frame_per_polyp}), which are important for realistic AI model benchmarking and training as outlined by recent literature \cite{nogueira-rodriguez_negative_2023, misawa2021development}. Furthermore, video frames from public datasets are seldom not at native spatial-temporal resolution and typically sourced from a limited number of centers. 

Physicians do not perform tasks such as polyp detection and classification in the real world by statically evaluating still or nearly perfect images or short video clips but instead via a process of temporal visual information reasoning \cite{cherubini_gorilla_2023, reverberi_experimental_2022}. The discrepancy between the available datasets and the real-world scenario inevitably affects both the design and development of CAD(e/x) algorithms, with a large part of academic research works to date still focusing on frame-by-frame approaches placing little emphasis on live processing speed and latency or full-procedure evaluation. Thus resulting in sub-optimal learning and unrealistic AI model performance assessments \cite{nogueira-rodriguez_negative_2023, misawa2021development, biffi_novel_2022, ali2024assessing}. Similarly, open research challenges\cite{bernal2017comparative,angermann2017towards,jha2020medico,hicks2021medico,hicks2021medai,ali2024assessing}, primarily centered on the accuracy of polyp detection, segmentation, and classification tasks, have gradually shifted their focus towards enhancing model robustness, speed and efficiency. However these challenges, employing datasets detailed in Table \ref{tab:1} and Table \ref{tab:2}, inherently reflect the above mentioned limitations by not fully capturing the dynamic and complex nature of real-world colonoscopy procedures.

\begin{table}[t!]
    \centering
    \scalebox{0.75}{%
    \begin{tabular}{|c|c|c|c|c|c|c|}
    \hline
        Dataset & Date of Publication & Format & Resolution & Annotation & Multiple Polyp Images & Non-Polyp Images \\ \hline
        CVC-ColonDB\cite{bernal2012towards} & 2012 & 300 images & 574$\times$500 & Segmentation & No & No  \\ \hline
        ETIS-Larib\cite{silva2014toward} & 2014 & 196 images & 1225$\times$966 & Segmentation & Yes & No  \\ \hline
        CVC-ClinicDB\cite{bernal2015wm} & 2015 & 612 images & 384$\times$288 & Segmentation & Yes & No  \\ \hline
        ASU-Mayo\cite{tajbakhsh2015automated} & 2015 & 18,781 images & unknown & Segmentation & No & Yes  \\ \hline
        CVC-ClinicVideoDB\cite{tajbakhsh2015automated} & 2017 & 11,954 images & 384$\times$288 & Segmentation & No & Yes  \\ \hline
        CVC-PolypHD\cite{angermann2017towards} & 2018 & 56 images  & 1920$\times$1080 & Segmentation & Yes & No  \\ \hline
        Kvasir-SEG\cite{jha2020kvasir} & 2020 & 1000 images & 1920$\times$1072 and 332$\times$487 & Segmentation & Yes & No  \\ \hline
        PICCOLO\cite{sanchez2020piccolo}  & 2020 & 3433 images  & 854$\times$480 and 1920$\times$1080 & Segmentation & Yes & Yes  \\ \hline
        KUMC\cite{li2021colonoscopy}   & 2021 & 37,899 images & various & Bounding Box & No & Yes  \\ \hline
        SUN\cite{misawa2021development}  & 2021 & 158,690 images & 1240$\times$1080 & Bounding Box & No & Yes  \\ \hline
        LDPolypVideo\cite{ma2021ldpolypvideo}  & 2021 & 40,187 images & 560x480 & Bounding Box & Yes & Yes  \\ \hline
        PolypGen\cite{ali2023multi}  & 2023 & 6282 images & various & Segmentation & Yes & Yes  \\ \hline
        REAL-Colon & 2023 & 2,757,723 images & 1920$\times$1080 & Bounding Box & Yes & Yes \\ \hline
    \end{tabular}
    }
    \caption{\label{tab:1} Available datasets for polyp detection for open research.  Datasets annotated with segmentation can also be utilized to derive bounding box annotations.}
\end{table}

\begin{table}[t!]
    \centering
    \scalebox{0.75}{%
    \begin{tabular}{|c|c|c|c|c|c|c|c|}
    \hline
        Dataset & Date of Publication & Format & Resolution & Histopathology & Size & Multiple Polyp & Non-Polyp  \\ \hline
        UAH\cite{mesejo2016computer} & 2016 & Videoclips with Video-Label & 768$\times$566 & SLL/Hyp/Ade & No & No & No \\ \hline
        PICCOLO\cite{sanchez2020piccolo}  & 2020 & Images with BB and Label  & 854$\times$480 and 1920$\times$1080 & SLL/Hyp/Ade & Yes & Yes & No  \\ \hline
        KUMC\cite{li2021colonoscopy}  & 2021 & Videoclips with BB and Label & various & Hyp/Ade & No & No & Yes \\ \hline
        SUN\cite{misawa2021development}  & 2021 & Videoclips with BB and Label & 1240$\times$1080 & Complete & Yes & No & Yes \\ \hline
        REAL-Colon & 2023 & Full-Procedure with BB and Label & 1920$\times$1080 & Complete & Yes & Yes & Yes \\ \hline
    \end{tabular}
    }
    \caption{\label{tab:2} Available datasets for polyp characterization and sizing for open research.}
\end{table}

This paper aims to bridge the gap between open and privately-funded research by introducing a comprehensive, multi-center dataset of unaltered, real-world colonoscopy videos. The REAL-Colon (Real-world multi-center Endoscopy Annotated video Library) dataset comprises recordings of sixty colonoscopies. In creating this dataset, a consortium comprising Sano Hospital in Hyogo, Japan; University Hospital of St. Pölten, Austria; and Ospedale Nuovo Regina Margherita in Rome, Italy, each provided a subset of fifteen patients. These patient cases were drawn from distinct clinical studies wherein colonoscopy procedures were documented on video as part of the study protocol.
Additionally, Cosmo Intelligent Medical Devices contributed by supplementing the dataset with fifteen patient cases from one of their sponsored studies conducted within the United States. Crucially, Cosmo Intelligent Medical Devices also annotated the entire dataset of 2.7M image frames to the highest quality standard. 

For each video, each colorectal polyp has been annotated with a bounding box in each video frame where it appeared by trained medical image annotations specialists, supervised by expert gastroenterologists to ensure their accuracy and consistency. Polyp information, including histology, size, and anatomical site, has been recorded, double-checked by annotation specialists and at least an experienced gastroenterologists, and reported with several other clinical variables. Patient variables obtained from electronic case report forms (eCRF), including endoscope brand and bowel cleanliness score (BBPS) have also been collected for each video. As illustrated in Tables \ref{tab:1} and \ref{tab:2}, the REAL-Colon dataset stands unparalleled in its scale, quality, and diversity, offering a singular asset for researchers and developers dedicated to pushing the boundaries of AI in colonoscopy. Furthermore, REAL-Colon uniquely enables comprehensive benchmarking of polyp detection algorithms against the backdrop of authentic, unedited full-procedure videos, markedly distinguishing our contribution from existing datasets.

\section*{Methods}

\subsection*{Data Cohorts}
The REAL-Colon dataset is a compilation of colonoscopy video recordings that combines diverse endoscopy practices across various geographical regions, thereby enhancing the heterogeneity of the physicians' maneuvers captured during the procedures. Each colonoscopy has been recorded in its entirety, from start to finish, at maximum resolution, devoid of any pauses or interruptions.
In the dataset, the first clinical study, denoted as "001", is a trial sponsored by Cosmo Intelligent Medical Devices. It pertains to the regulatory approval of an AI medical device platform, GI Genius\texttrademark, in the United States. This randomized controlled trial (identified as NCT03954548 in ClinicalTrials.gov) was conducted from February 2020 to May 2021 in Italy, the United Kingdom, and the United States\cite{wallace_impact_2022}. However, only patients from the three participating U.S. clinical centers were included in the REAL-Colon dataset.
The second clinical study, tagged as "002" in the dataset, is a single-center, prospective non-profit study (identifier NCT04884581 in ClinicalTrials.gov) conducted in Italy from May 2021 to July 2021\cite{hassan_artificial_2022}.
In addition to these structured pre-registered studies, two clinical centers contributed video-recorded colonoscopies from their internal acquisition campaigns designed for scientific research. The first campaign, termed "003" in the dataset, recruited patients from the University Hospital of St. Pölten, Austria's Gastroenterology and Hepatology and Rheumatology department. The second campaign, labeled "004", involved patients from the Gastrointestinal Center, Sano Hospital, Hyogo, Japan.

Before participating, all patients from the four data acquisition efforts provided written consent for their anonymised data to be used in research studies. The two clinical studies ("001" and "002") and the two acquisition campaigns ("003" and "004") each received the necessary approvals from their respective Ethical Committees or local Institutional Review Boards. These were as follows: "001" - Mayo Clinic (approval number: 19-007492), "002" - Comitato Etico Lazio 1 (611/CE Lazio1), "003" - Ethikkommission Niederösterreich (GS1-EK-3/196-2021), "004" - Sano Hospital (202209-02). All relevant patient data during the acquisition were recorded via an electronic case report form (eCRF) system. 

The participants from all four cohorts were patients aged 40 years or above undergoing colonoscopy for primary CRC screening, post-polypectomy surveillance, positive fecal immunochemical tests, or symptom/sign-based diagnosis. Exclusion criteria included a history of CRC or inflammatory bowel disease, a history of colonic resection, emergency colonoscopy, or ongoing antithrombotic therapy.
Standard resection techniques were utilized throughout each colonoscopy procedure to excise detected polyps. Endoscopists documented various polyp characteristics, including size, anatomical location, and morphological traits, in accordance with the Paris classification\cite{participants_in_the_paris_workshop_paris_2003}. Subsequently, each polyp, whether resected or biopsied, underwent expert pathological analysis. Resected tissue samples confirmed histologically as colorectal polyps were classified per the Revised Vienna Classification\cite{schlemper_vienna_2000}.
By integrating data from these four diverse cohorts across six medical centers spanning three continents, we have strived to build a comprehensive, heterogeneous dataset that robustly represents real-world colonoscopy practices.

\subsection*{Video recording}
The video acquisition process was executed to preserve the quality of the original footage and was the same for the four cohorts. Professional video recorders, capable of capturing YUV video streams with 4:2:2 chroma subsampling and 10-bit depth, were used to ensure no loss in color or resolution. These recorders effectively captured the original streams at a resolution of 1920x1080 (interlaced), resulting in high-definition video material. Standard endoscopy equipment manufactured by Olympus and Fujifilm was used. Following the recording, videos were compressed using the high-quality Apple ProRes codec to ensure the preservation of the image quality while reducing the overall file size. The videos were subsequently converted into individual frames using the ffmpeg software tool, opting for JPEG to balance image quality with file size, thereby managing the overall dataset size. This conversion was carried by configuring the -qscale:v, -qmin, and -qmax options to 1, thus minimizing compression impact while avoiding the large file sizes associated with lossless formats.

\subsection*{Anonymization protocol}
We established a full-anonymization protocol before transferring videos and clinical data from individual studies to Cosmo Intelligent Medical Devices for dataset labeling and compilation. This protocol entailed the removal of all direct identifiers, including personal contact details. Specifically, videos were edited at source to eliminate any on-screen data from the Electronic Medical Records (EMR) local system, ensuring that frames were free from any information that could reveal patient or procedure identities. Additionally, all direct identifiers were removed at source from electronic Case Report Forms (eCRFs) and all dataset frames cropped to the endoscope's field of view. Consequently, only demographic quasi-identifiers such as age and sex were shared with Cosmo Intelligent Medical Devices.  This full-anonymization protocol, supported by stringent security and privacy measures, effectively reduced data stewardship obligations under GDPR and HIPAA terms. 

\begin{figure}[t!]
\centering
\includegraphics[width=\linewidth]{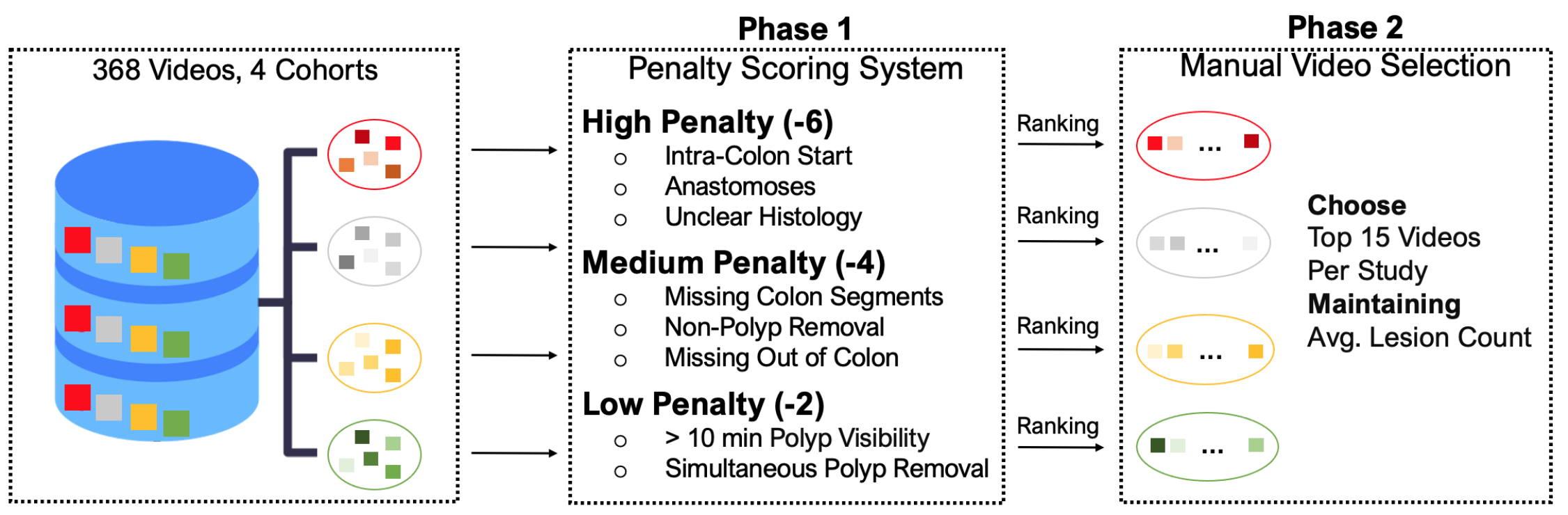}
\caption{Flowchart outlining the two-phase selection process for creating the REAL-Colon dataset from 368 video recordings across four distrinct cohorts. Phase 1 applies a penalty scoring system based on video and histological criteria, leading to Phase 2, where the 15 videos per cohort are manually selected, after ranking, to ensure diversity and representation while maintaining the cohort average lesion count.}
\label{fig:flowchart}
\end{figure}

\subsection*{Dataset design}
The selection of our dataset from an initial pool of 368 video recordings across four distinct cohorts aimed to create a diverse and representative dataset. To achieve this, we implemented a two-step strategy to curate a set of 60 videos, with fifteen videos independently sampled from each cohort, as depicted in Figure \ref{fig:flowchart}.

Our method was grounded in a penalty scoring system in the first selection phase (Figure \ref{fig:flowchart}, Phase 1). We established three penalty categories: high (-6), medium (-4), and low (-2). Each penalty point was cumulative and calculated based on the deviation from ideal recording conditions. Low penalties were given to videos showing polyps visible for more than ten minutes, as such instances would skew the video duration towards long resection maneuvers. Simultaneous polyp removal also attracted a low penalty. This situation, where more than one polyp is resected with the same instrument, can lead to ambiguous associations between polyps shown in the video recordings and histological descriptions of polyps.

Medium penalties were assigned to videos that omitted out-of-colon sequences before and after the procedure. These segments of the video are important as AI algorithms must operate effectively beyond the colon, and their omission could potentially bias the recordings. Situations where all colon sections were not identifiable during withdrawal or when non-polyp biopsies and polypectomies were performed also received a medium penalty.

Videos that initiated directly within the colon, displayed anastomoses, or did not have available histological information for all excised lesions, were subjected to high penalties. Importantly, high penalty points were given when non-polyps (lymphoid follicles, lymphoid aggregates, ulcers, lipomas, or healthy tissue) histological results were unclear. This could imply random biopsies, incorrect resections, or ambiguous histology, all of which could lead to confusion or misinterpretation of the data.

During the second phase (Figure \ref{fig:flowchart}, Phase 2), our selection favored videos with the least penalties. We balanced this with the need to maintain a representative dataset. For this reason, even among the videos with the highest penalties, we ensured the average lesion count within each selected cohort approximated the total dataset's average. This procedure preserved the distribution of lesions in our final selection, mirroring the original dataset. The highest penalty within our curated REAL-Colon dataset was -4.

Given their critical role in CAD model training, no specific criteria were initially set for histology ground truth classifications (adenoma, non-adenoma). However, recognizing their importance, we conducted a retrospective review and found that the distribution of these classifications was in line with the originating clinical studies, confirming the validity of our approach.

\subsection*{Clinical Data}
The REAL-Colon dataset is structured such that each video corresponds to an individual patient, and may or may not contain instances of colon polyps. As such, there are clinical values that pertain to the patient and others that are specific to the polyps. Clinical information associated with each video includes patient demographics such as age and sex and technical details of the endoscope used, including the brand, the frames per second (fps) of the video, and the number of polyps resected during the procedure.

An essential quality parameter at the patient level is the evaluation of the colon's cleanliness. Before a colonoscopy, patients are administered a laxative to eliminate solid waste from the digestive tract swiftly. The effectiveness of this preparation procedure is routinely evaluated using a clinical scale known as the Boston Bowel Preparation Scale (BBPS) \cite{dekker_advances_2018}. This scale, utilized by endoscopists and ranging from 0 to 9, provides a measure of the quality of bowel preparation following cleansing procedures, thereby enhancing the clinical relevance of the dataset.

The dataset encompasses a broad range of patients, with an average age of $64.6 \pm 10.7$ and a male predominance of $55$\%. Notably, 14 out of the 60 videos (representing 23\%) within the dataset do not contain any polyps, which underscores the variety of clinical scenarios captured.

\begin{figure}[t!]
\centering
\includegraphics[width=\linewidth]{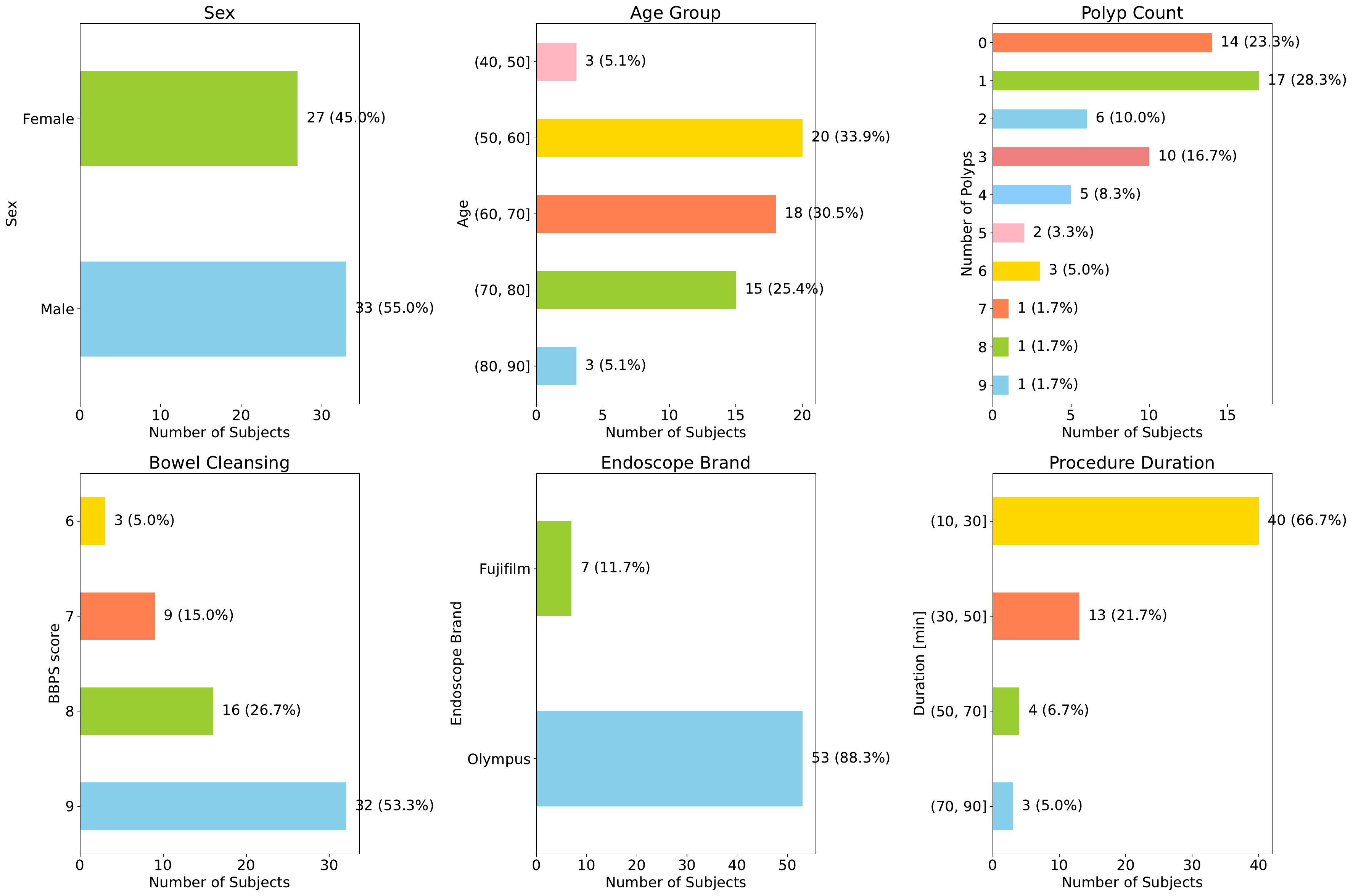}
\caption{Clinical Data Distribution. This figure presents histograms depicting the distribution of sex, age, polyp count per procedure, BBPS scores, endoscope brand, and procedure duration within the REAL-Colon dataset.}
\label{fig:clinical_data}
\end{figure}

For a comprehensive understanding and visualization of the distribution of these variables within our dataset, we direct readers to Figure \ref{fig:clinical_data}. This figure presents histograms detailing these parameters, thereby providing a more holistic view of the dataset.

\subsection*{Polyp Detection Annotation}
A team of ten medical image annotation specialists executed a comprehensive bounding box annotation protocol under the supervision of expert gastroenterologists. A key aspect of the protocol involved annotating the polyp, beginning from its total resection and proceeding in reverse frame by frame until its initial appearance. This approach facilitated lesion tracking across sequential frames, even when the lesions were temporarily invisible. 

To accomplish the task, the team employed a specialized, in-house annotation tool from Cosmo Intelligent Medical Devices. Although not available for public use, this tool mirrors the functionalities of many other video annotation software, enabling the sequential identification of polyps and the application of bounding boxes around them within each frame.

To ensure accuracy and consistency, the polyp annotation process was iterative, with annotations undergoing refinements through weekly meetings. These sessions provided a platform for collaboration, promoting valuable exchanges between the annotation team and the supervising expert gastroenterologists.

This procedure culminated in a dataset comprising 2,757,723 frames and 132 excised colorectal polyps. The process yielded a total of 351,264 bounding box annotations.  

\begin{figure}[t!]
\centering
\includegraphics[width=\linewidth]{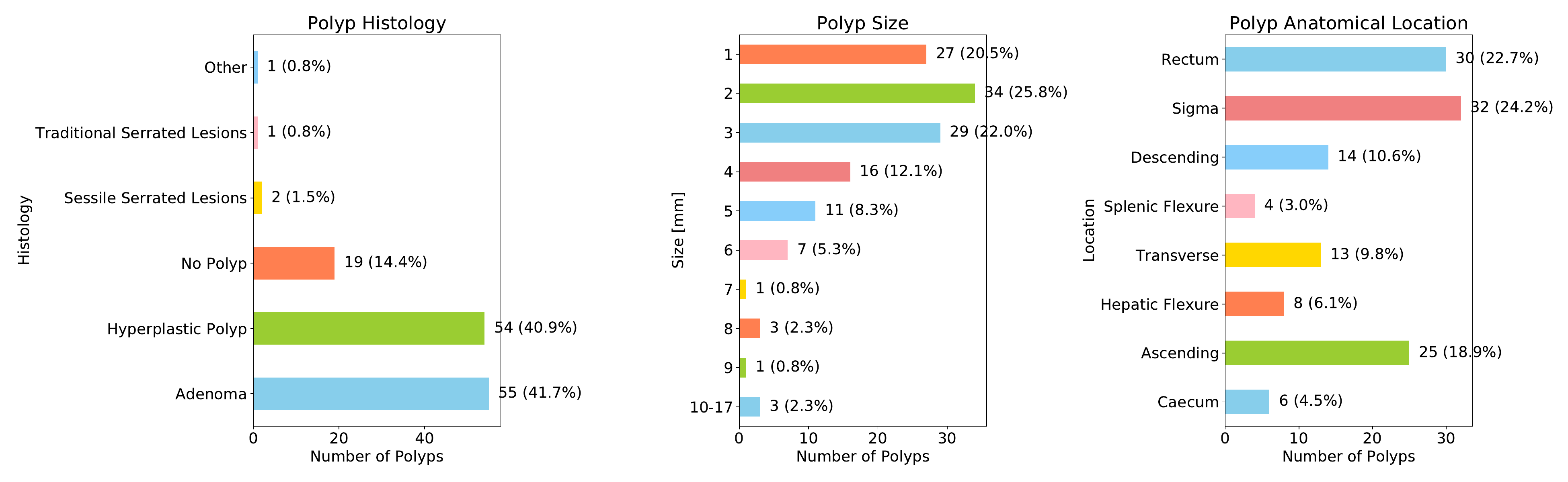}
\caption{Polyp Characteristics Distribution. The histograms in this figure highlight the distribution of the anatomical location, size (in millimeters), and histology of the polyps included in the REAL-Colon dataset.}
\label{fig:polyp_data}
\end{figure}

\subsection*{Polyp Histopathological Information}
Each annotation of polyps underwent cross-verification against its corresponding eCRF for each patient. The objective of this validation process, supervised by an expert gastroenterologist, was to eliminate any potential discrepancies between the video annotations and the corresponding histological data. 

The dataset incorporated detailed information for each polyp, including size, colon anatomical location, and histopathological results. The distribution of this data is represented by histograms in Figure \ref{fig:polyp_data}. Notably, the histological analysis indicated that out of 132 resected polyps, 19 (14\%) were not histologically classified as polyps. These were identified as lymphoid follicles, lymphoid aggregates, ulcers, lipomas, or healthy tissue. 

Moreover, the dataset presents comprehensive histological data and maintains a balanced distribution between adenomatous and non-adenomatous polyps, constituting 40\% each. One polyp was classified as "other," which was identified as a fibrous anal polyp.

\section*{Data Records}
The complete dataset is now available for download on Figshare, as referenced\cite{biffi_2023_real}. It has been made available under the Creative Commons Attribution (CC BY) license, facilitating both educational and research applications. Users are encouraged to acknowledge this paper when utilizing the dataset in their work.

Bounding box annotations for each ground-truth polyp detection in the dataset were stored in the MS COCO format \cite{lin2014microsoft}. An XML file was generated for each frame, following the MS COCO template. The dataset includes the following components:

\begin{itemize}
    \item 60 compressed folders titled \texttt{SSS-VVV\_frames}, each containing frames from a specific recording.
    \item 60 compressed folders titled \texttt{SSS-VVV\_annotations}, each encompassing the annotations for each frame.
    \item A \texttt{video\_info.csv} file, providing metadata for each video. This includes unique video ID, video FPS, number of frames, original cohort (from 001 to 004), patient's age and sex, number of polyps discovered in the video, BBPS score, and the brand of the endoscope used.
    \item A \texttt{lesion\_info.csv} file, offering metadata for each polyp, such as unique polyp ID, the unique video ID it belongs to, polyp size (in millimeters), the polyp's colon location, histology, and extended histology, which presents additional information from the histological report in the eCRF.
    \item A \texttt{dataset\_description.md} file, which is a README file providing an overview of the dataset.
\end{itemize}

\section*{Technical Validation}

\subsection*{Polyp Dynamics and Characteristics}
\begin{figure}[t!]
\centering
\includegraphics[width=\linewidth]{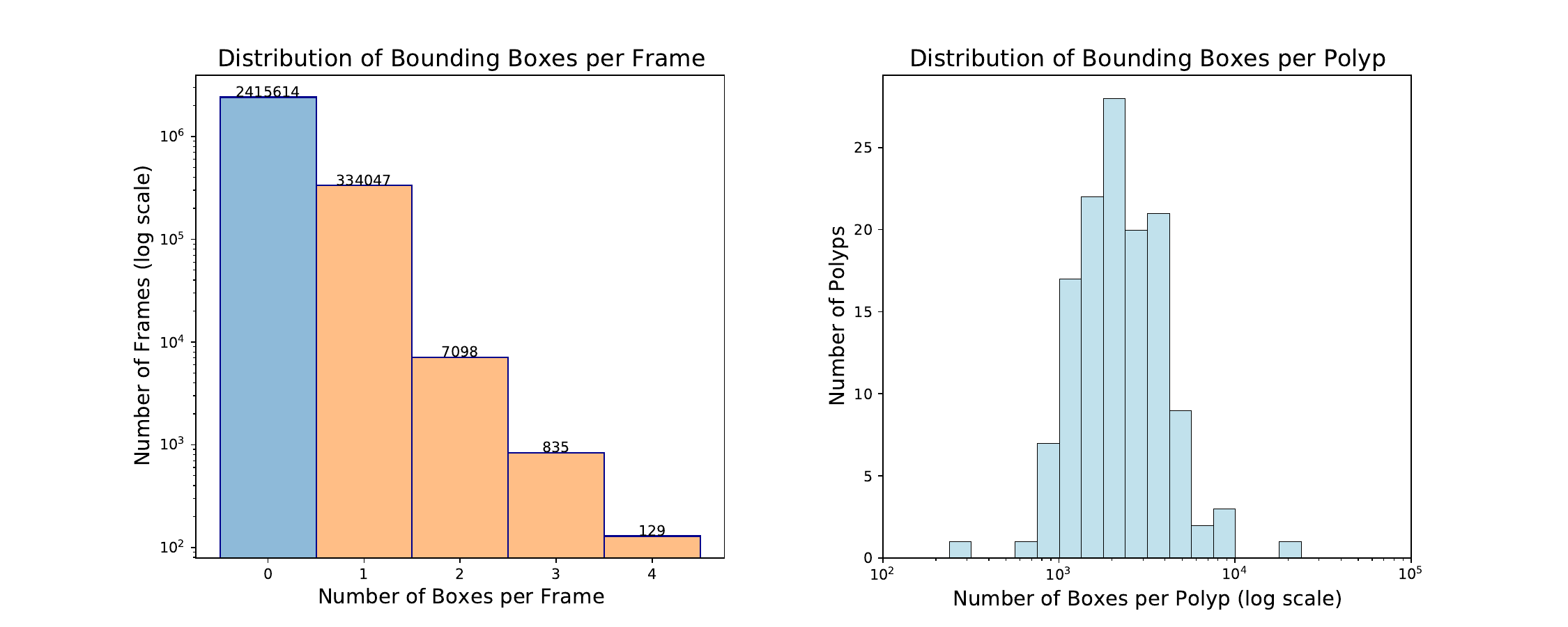}
\caption{Left, a histogram displaying the number of boxes per frame. On the right, the distribution of the number of bounding boxes associated to each polyp.}
\label{fig:hist_boxes_per_frame_per_polyp}
\end{figure}

Figure \ref{fig:hist_boxes_per_frame_per_polyp} (left) displays the distribution of bounding boxes per frame in the REAL-Colon dataset\cite{biffi_2023_real}. Of 2,757,723 total frames, 87.6\% (2,415,614) do not contain any bounding box annotation, while 12.4\% (342,109) feature at least one from the 132 excised colorectal polyps. Notably, less than 0.3\% of frames contain multiple polyps - 2.7\% of positive frames - peaking at four. When considering only the frames within the first 5 seconds of polyp appearance, the occurrence of multiple polyps increases to 7.9\%. This is relevant prior information for learning-based computer vision algorithms and highlights the distinct nature of detection, tracking, and classification tasks in colonoscopy compared to standard computer vision tasks that often involve simultaneous detection of numerous classes or objects, as seen in MS COCO or Kinetics datasets.

Figure \ref{fig:hist_boxes_per_frame_per_polyp} (right) illustrates the distribution of bounding boxes per polyp, indicating a mean appearance duration of 1 minute and 15 seconds per polyp, with the longest enduring over 12 minutes.

\begin{figure}[t!]
\centering
\includegraphics[width=0.9\linewidth]{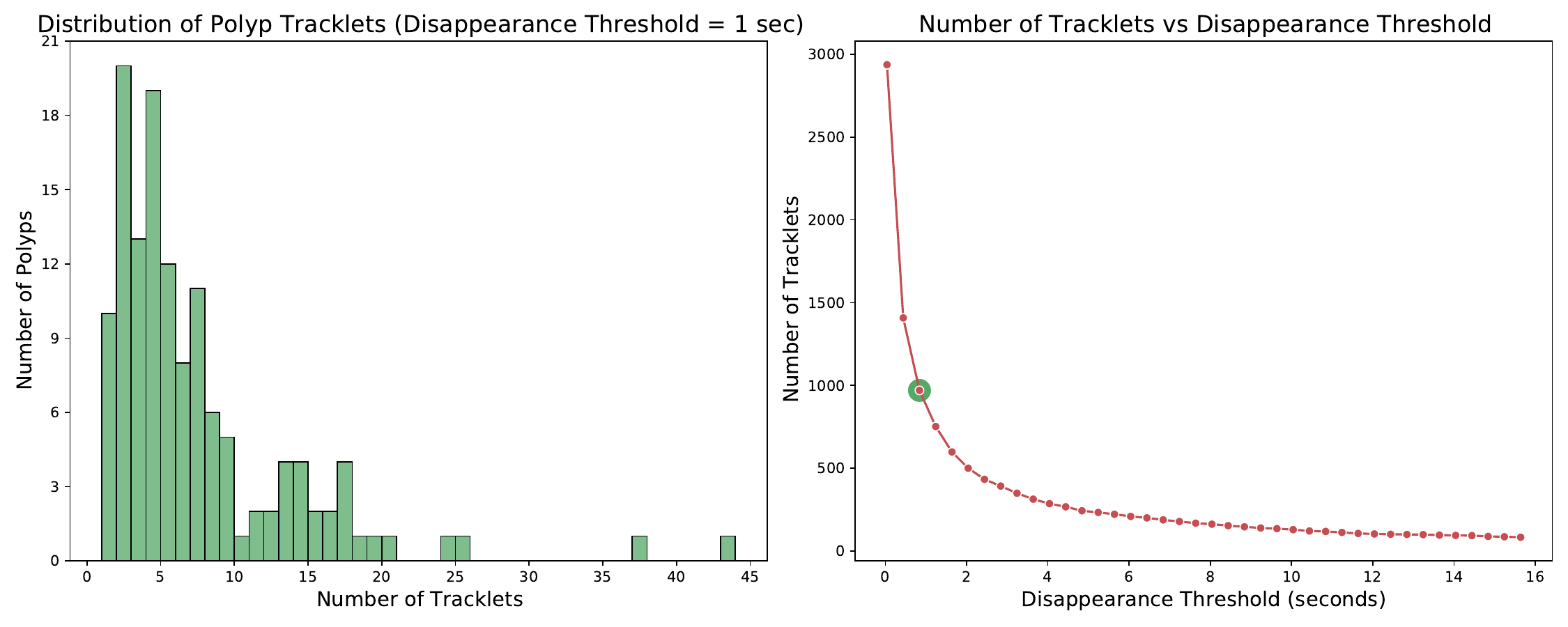}
\caption{Left: Histogram displaying the number of tracklets per polyp, using a 1-second threshold to identify separate tracklets. The x-axis represents the number of tracklets associated with each polyp, while the y-axis shows the count of polyps with that number of tracklets. Right: Plot illustrating the decrease in the number of tracklets as a function of the disappearance threshold. Here, the x-axis signifies the disappearance threshold in seconds, which determines when a new tracklet is created once a polyp disappears for longer than the threshold duration. The y-axis reports the resulting number of tracklets.}
\label{fig:tracklets_plots}
\end{figure}

The bounding box detections corresponding to a single polyp are not necessarily contiguous. They may fracture into numerous tracklets as polyps can momentarily disappear from view. This phenomenon is elucidated on the left side of Figure \ref{fig:tracklets_plots}, which showcases a histogram of the number of tracklets into which the dataset polyps are divided. In this figure, the criterion for identifying a new tracklet is the absence of a bounding box instance for that polyp for more than one second. It can be noticed that utilizing this criterion, only 10 out of 132 polyps have a continuous appearance without disappearing from view for more than 1 second. This underlines the relevance of the task of polyp tracking for real-time colonoscopy applications where accurately re-identify tracklets is essential to prevent the need to restart temporal analysis with each polyp disappearance. Furthermore, the right side of Figure \ref{fig:tracklets_plots} presents the total number of tracklets that emerge when applying different temporal thresholds. For instance, when applying a 15-second threshold, over 100 tracklets persist, possibly signifying the most challenging cases.

\begin{figure}[t!]
\centering
\includegraphics[width=\linewidth]{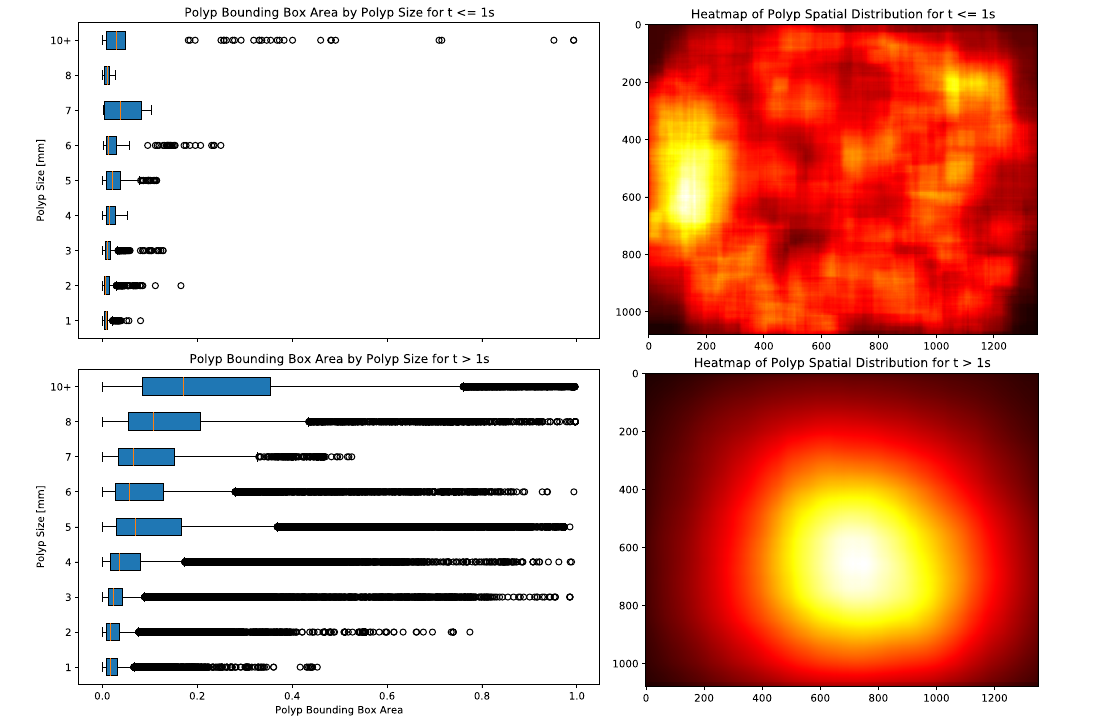}
\caption{Boxplots contrasting actual polyp sizes with bounding box dimensions (left) and heatmaps depicting bounding box placements (right) during the early phase of appearance ($\leq$ 1s) and afterwards (> 1s). In the early frames, polyps are captured within small bounding boxes scattered across the colon. As time progresses, the endoscopist centralizes the polyps in the frame, leading to larger and more variable in dimensions bounding boxes.}
\label{fig:heat_boxplot}
\end{figure}

Figure \ref{fig:heat_boxplot} (right) presents two heatmaps illustrating the spatial distribution of polyps within the frames, stratified by their time of appearance: the initial second and thereafter. The heatmaps emphasize that endoscopists tend to frame polyps closer to the center of the image as time progresses, despite their initial scattered appearance. In contrast, Figure \ref{fig:heat_boxplot} (left) exhibits two boxplots contrasting the bounding box dimensions over time with the actual sizes (in millimeters) as indicated in the dataset. These plots reveal that irrespective of their actual sizes, all bounding boxes initially have smaller dimensions during early detection.  Additionally, it can be observed that as detection progresses, all polyps are characterized by numerous bounding boxes occupying more than 50\% of the image.  This indicates that determining polyp size solely based on bounding box dimensions is challenging and requires additional reasoning on contextual spatial-temporal information.

\subsection*{Polyp Detection}
This section presents a technical validation conducted to evaluate polyp detection in the REAL-Colon dataset, primarily focusing on verifying the quality and usefulness of the dataset. For this task, we leveraged an off-the-shelf SSD object detection model \cite{liu2016ssd}, as implemented in the NVIDIA's DeepLearningExamples repository. Comprehensive training and testing code is accessible via our GitHub repositories at 
\href{https://github.com/cosmoimd/DeepLearningExamples}{https://github.com/cosmoimd/DeepLearningExamples}.

\begin{table}[t!]
    \centering
    \scalebox{0.8}{%
    \begin{tabular}{|c|c|c|c|c|c|c|c|c|c|c|}
    \hline
        Num Frames per Lesion & Ratio of Neg Frames & \#Train Images  & AP & AP$_{50}$ & AP$_{75}$ & FPR & TPR \\ \hline
        100 & 0 & 11141  & 0.107 & 0.181 & 0.117 & 0.123 & 0.463  \\ \hline
        1000 & 0 & 107525 & 0.137 & 0.231 & 0.147 & 0.082 & 0.464 \\ \hline
        all & 0 & 230936 & 0.165 & 0.272 & 0.181 & 0.080 & 0.495 \\ \hline
        all & 0.01 & 246556 & 0.190 & 0.286 & 0.217 & 0.058 & 0.434 \\ \hline
        all & 0.05 & 309116 & 0.193 & 0.311 & 0.219 & 0.072 & 0.472\\ \hline
        all & 0.1 & 387315 & 0.178 & 0.265 & 0.205 & 0.083 & 0.454 \\ \hline
        all & all & 1794907 & $\mathbf{0.216}$ & $\mathbf{0.338}$  & $\mathbf{0.245}$ & $\mathbf{0.054}$ & $\mathbf{0.505}$ \\ \hline
    \end{tabular}}
    \caption{\label{tab:validation} This table presents detection results from experiments utilizing different training dataset subsets, with metrics aligned to COCO object detection.  AP measures the mean Average Precision across a range of IoU thresholds from 0.5 to 0.95, while AP${50}$ and AP${75}$ specifically refer to IoU thresholds of 0.5 and 0.75, respectively. The False Positive Rate (FPR) and True Positive Rate (TPR) indicate the percentage of polyp-negative frames in which models erroneously detected bounding boxes and the percentage of positive frames in which models flagged an alert, respectively. Testing was conducted on the entire 591,647 frames test set. Each row corresponds to a distinct subset of training images from the 40 videos in the training split, varying the number of selected positive frames per lesion and the ratio of selected negative frames to total negative frames per video.}
\end{table}

\begin{table}[t!]
    \centering
    \scalebox{0.8}{%
    \begin{tabular}{|c|c|c|c|c|}
    \hline
        Polyp & \#Test Images  & \#Test Polyps & AP & AR \\ \hline
    \hline
        Diminutive & 52,133 & 17 & 0.347 & 0.450  \\ \hline
        Non-Diminutive  & 29,973 & 4 & 0.120 & 0.260   \\ \hline
    \hline
        Adenoma & 37,264 & 12& 0.427 &  0.498  \\ \hline
        Non-Adenoma & 40,525 & 8& 0.145 &  0.295 \\ \hline
    \hline
        Hyperplastic Sigma-Rectum & 10,217 & 5 & 0.387 & 0.505  \\ \hline
        Others & 71,889 &16& 0.260 & 0.364 \\ \hline
    \hline
        All & 82,106 & 21 & 0.277  & 0.381 \\ \hline
    \end{tabular}}
    \caption{\label{tab:validation2}Average Precision and Average Recall (AR) for IoU=0.5 to 0.95 on different subsets of positive polyp frames within the test set.  The data is segmented to detail performance distinctions among different subsets of the training set alongside a comprehensive assessment for all the polyp positive test images.
}
\end{table}

Our experiments do not merely target the absolute performance of the proposed method on the REAL-Colon dataset; instead, we focus on evaluating the dataset's utility for this specific task. Specifically, we aimed to ascertain the influence of possessing a substantial sample set for each polyp on the accuracy of the trained model. Moreover, we conducted multiple training experiments incorporating varying proportions of negative frames from each video (1\%, 5\%, 10 \% and 100 \%  of negative frames from each video) in the training dataset to evaluate their impact on model performance.

We partitioned the first 10 videos from each cohort into the training set, the following two videos into the validation set, and the remaining three videos into the test set. For each model training session, we used a batch size of 96 and an image resolution of 300x300. Standard augmentation techniques, including scaling, random cropping, horizontal flipping, and image normalization, were applied, using the same parameters as those utilized for the MS COCO Dataset in the repository. All models were trained and tested on an Nvidia Tesla V100 GPU.

In Table \ref{tab:validation}, we report performance in terms of Average Precision (AP), measuring the mean AP across a range of IoU thresholds from 0.5 to 0.95 (measuring overlap between model bounding box predictions and GTs), in steps of 0.05. AP50 and AP75, instead, specifically refer to the model Average Precision at IoU thresholds of 0.5 and 0.75, respectively, and indicate more accurate detection performance. For every model, we also the False Positive Rate (FPR) and the True Positive Rate (TPR) per video, defined as the percentage of negative frames in which models erroneously detected bounding boxes, indicating the rate of false alarms on negative frames, and the percentage of positive frames in which at least a bounding box was detected. Throughout the training process, the validation set was used to monitor the AP at various epochs, and the best-performing model on the validation set throughout the training was selected. All experiments were tested on the same test set of 591,647 frames, including all frames (with and without boxes) from 12 test videos.

\begin{table}[t!]
    \centering
    \scalebox{0.8}{%
    \begin{tabular}{|c|c|c|c|c|c|}
    \hline
        Appearance & \#Test Images  & AP & AR \\ \hline
        $<1s$ & 570  & 0.138 & 0.254 \\ \hline
        $<3s$ & 1,710 & 0.179 & 0.334\\ \hline
        All & 82,106 & 0.217 & 0.381 \\ \hline
    \end{tabular}}
    \caption{\label{tab:validation3}Performance of the best SSD detection model on frames from the initial second and first three seconds of polyp appearance, rather than evaluating all positive frames.}
\end{table}

The initial analysis presented in Table \ref{tab:validation} highlights that incorporating a greater number of positive samples, despite them representing identical polyps, significantly boosts the model's performance. Subsequent rows detail the effects of varying negative sample proportions in the training set, indicating optimal SSD model performance with the full inclusion of available negative images alongside all positive frames. Although the current strategy prioritizes this comprehensive approach, future advancements might benefit from a predefined negative-to-positive frame ratio per training epoch to refine the training process further. The FPR and TPR results also underscore accuracy improvements by fully incorporating negative frames during training. The FPR is a critical metric because false alerts can lead to operator fatigue and distractions. Minimizing FPR throughout the entire procedure, while maintaining a high TPR, is essential to develop robust CADe systems. The REAL-Colon dataset enables such evaluations, facilitating the optimization of these key performance indicators. 

In Table \ref{tab:validation2}, we present the performance of the best model from Table \ref{tab:validation} (last row), detailing AP (Average Precision) and Average Recall (AR) across a spectrum of IoU thresholds from 0.5 to 0.95. This analysis encompasses different subsets of positive polyp frames within the test set, specifically distinguishing between frames containing diminutive or non-diminutive polyps, adenomatous polyps (including adenoma and traditional serrated adenoma (TSA) polyps) versus non-adenomatous polyps (including polyps with sessile serrated lesion (SSL) and hyperplastic (HP) histology), and excluding polyps not falling into these two sub-categories. Additionally, by integrating polyp anatomical location information with histology data, we were able to compute performance metrics for hyperplastic polyps in the sigmoid-rectum compared to all other locations. This distinction is crucial because hyperplastic polyps in the sigmoid-rectum exhibit different biological behaviors and cancer risk profiles compared to those in other parts of the colon, representing an example of the importance of location-specific performance evaluation for more precise and clinically relevant AI model assessments. Ideally, a robust model should demonstrate uniformly high performance across these varied classifications. Future efforts should concentrate on exploring augmentation techniques, algorithmic modifications, and training strategies to ensure such robust performance across all categories.

In Table \ref{tab:validation2}, we present the performance of the top-performing model from Table \ref{tab:validation}, specifically focusing on frames from the initial second and first three seconds of polyp appearance, rather than evaluating all positive frames. This early detection window is critical, and the REAL-Colon dataset facilitates such an analysis due to its frame-by-frame annotation at full temporal resolution. The results highlight the challenges of early detection across all models evaluated in our study, underscoring the difficulty in accurately identifying polyps during these initial moments. We advocate for dedicated research efforts to further enhance model performance during these crucial early stages of polyp appearance.

\begin{figure}[t!]
  \centering
  \begin{subfigure}[b]{0.24\textwidth}
    \includegraphics[width=\textwidth]{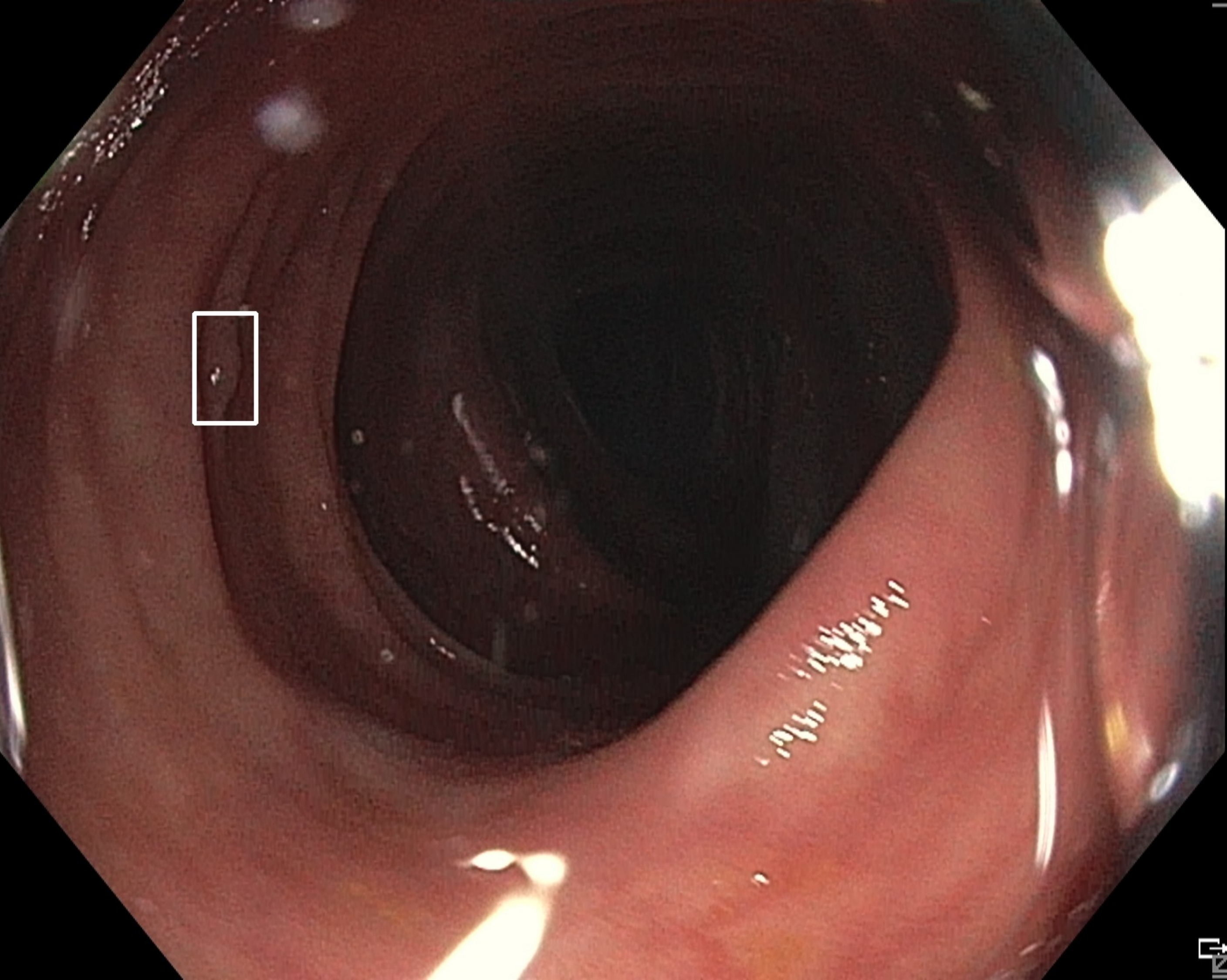}
    \caption{}
        \label{sfig:1}
  \end{subfigure}
  \hfill
  \begin{subfigure}[b]{0.24\textwidth}
    \includegraphics[width=\textwidth]{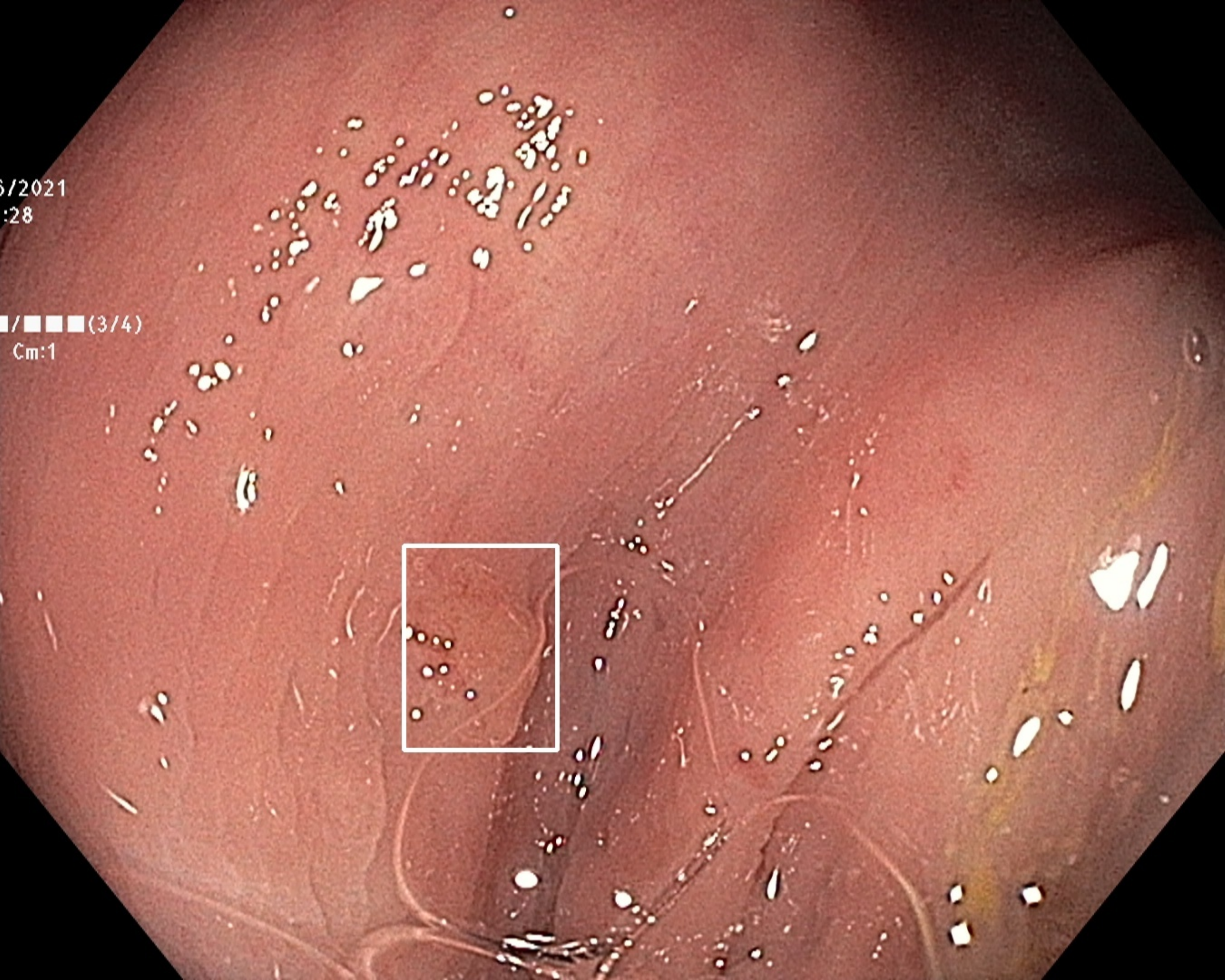}
    \caption{}
        \label{sfig:2}
  \end{subfigure}
  \hfill
  \begin{subfigure}[b]{0.24\textwidth}
    \includegraphics[width=\textwidth]{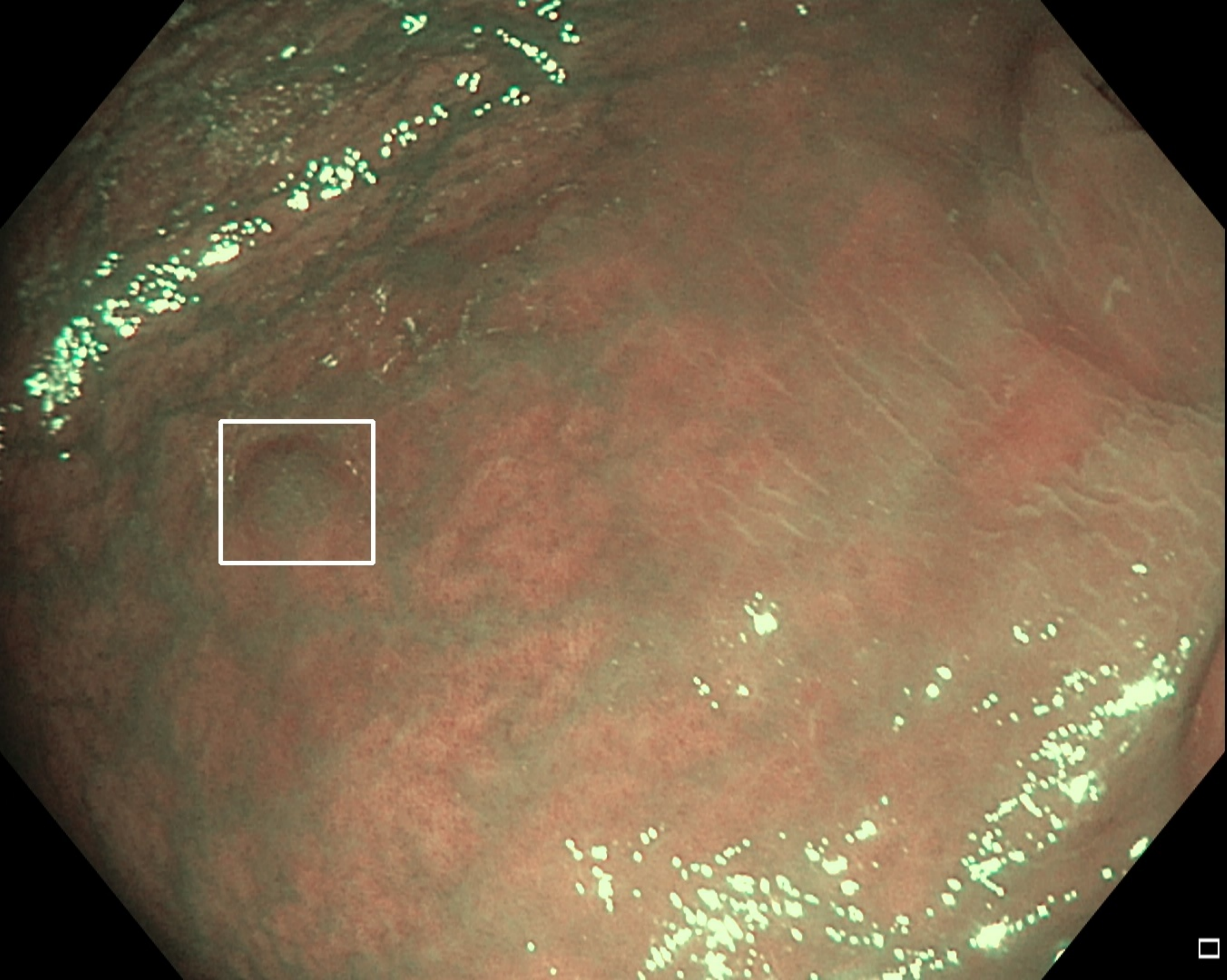}
    \caption{}
        \label{sfig:3}
  \end{subfigure}
  \hfill
  \begin{subfigure}[b]{0.24\textwidth}
    \includegraphics[width=\textwidth]{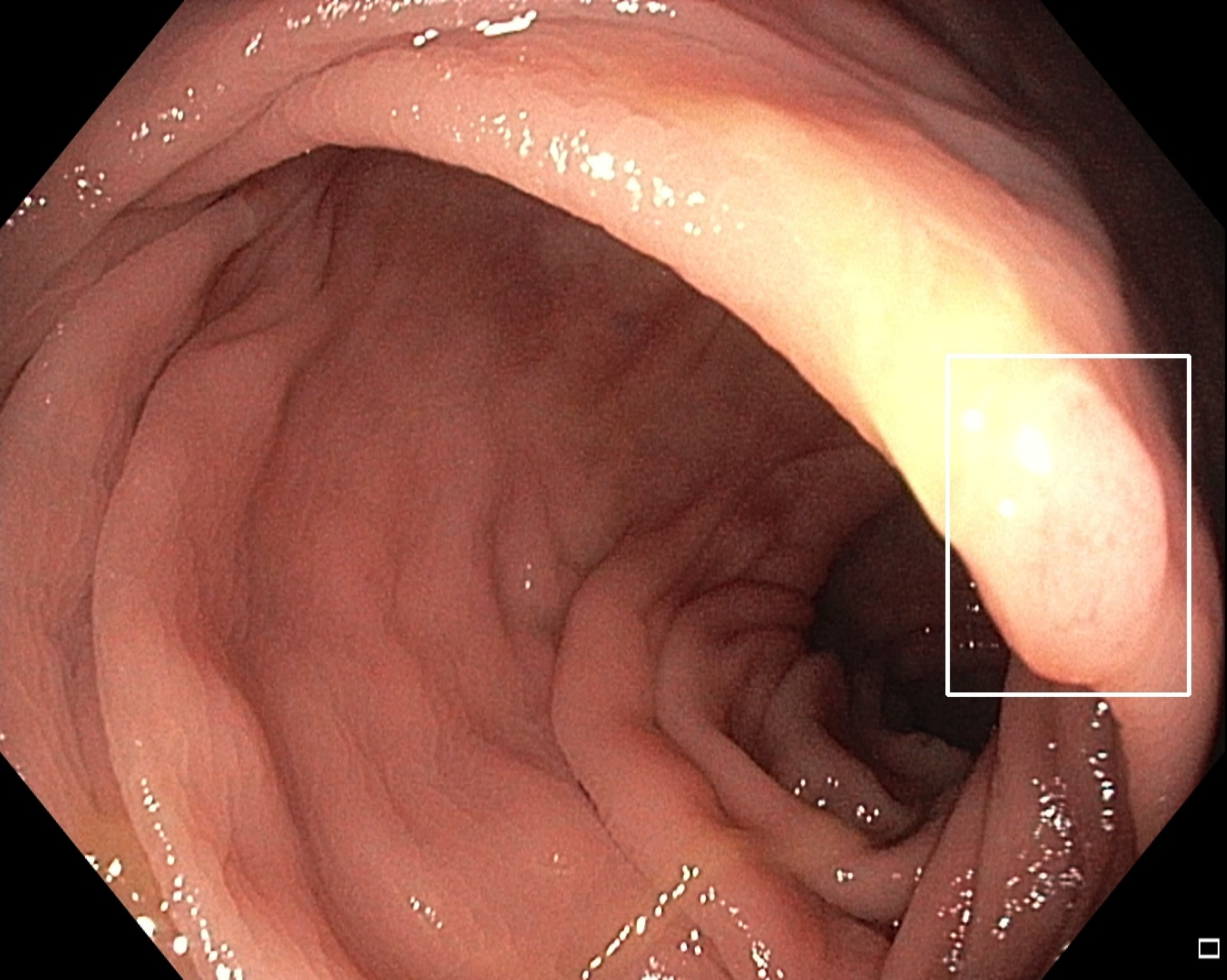}
    \caption{}
        \label{sfig:4}
  \end{subfigure}
  \vspace{2pt}
  
  \begin{subfigure}[b]{0.24\textwidth}
    \includegraphics[width=\textwidth]{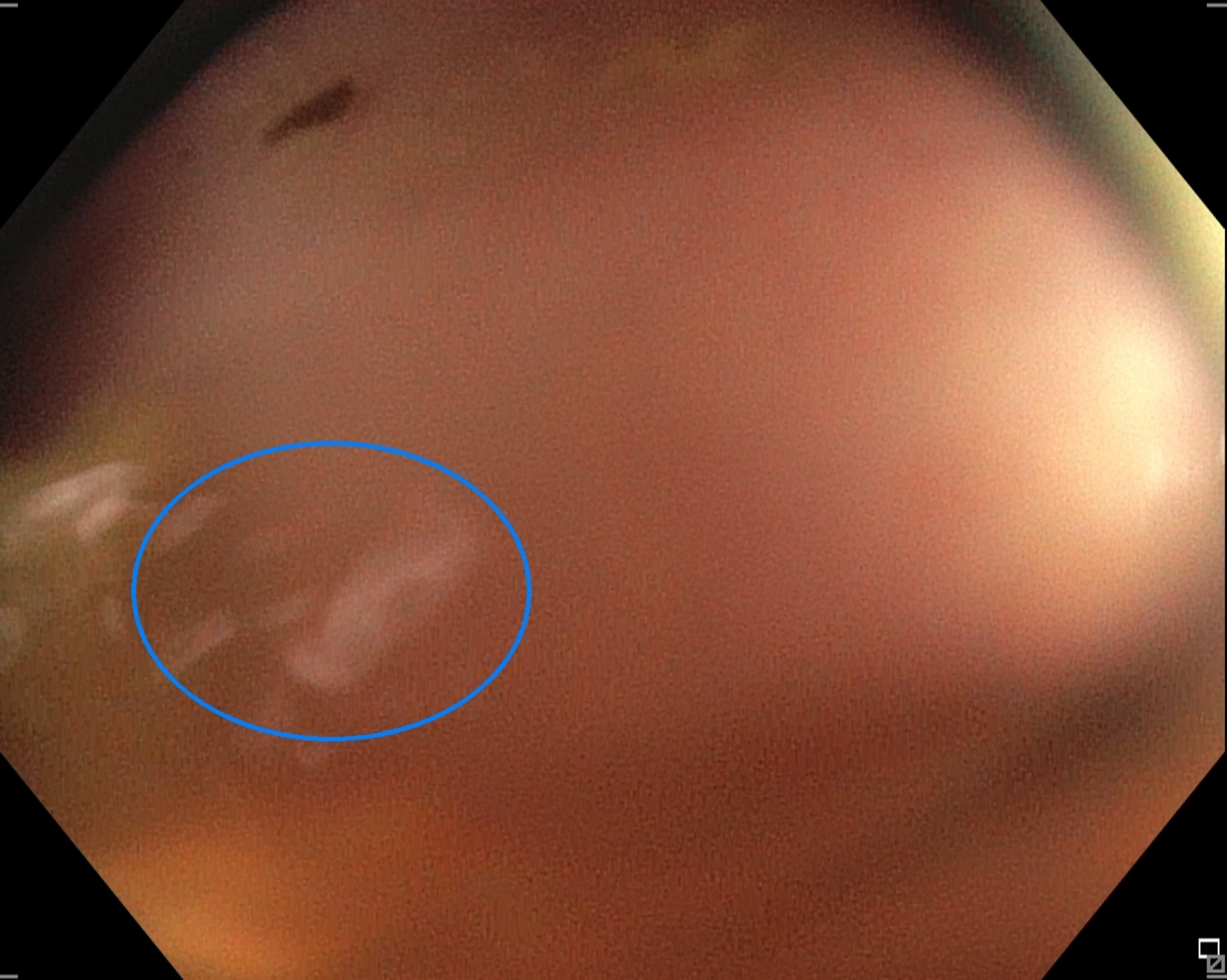}
    \caption{}
        \label{sfig:5}
  \end{subfigure}
  \hfill
  \begin{subfigure}[b]{0.24\textwidth}
    \includegraphics[width=\textwidth]{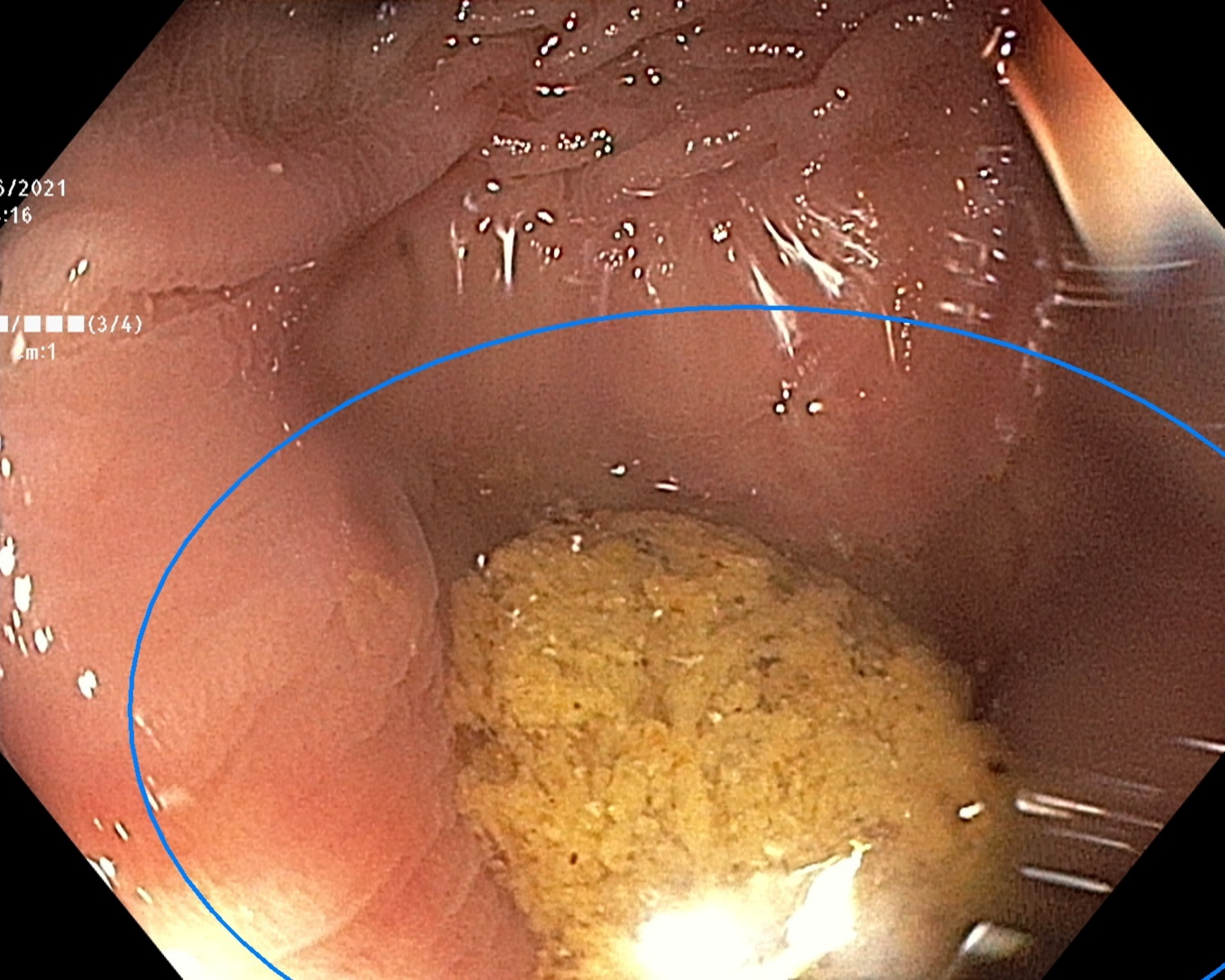}
    \caption{}    
    \label{sfig:6}
  \end{subfigure}
  \hfill
  \begin{subfigure}[b]{0.24\textwidth}
    \includegraphics[width=\textwidth]{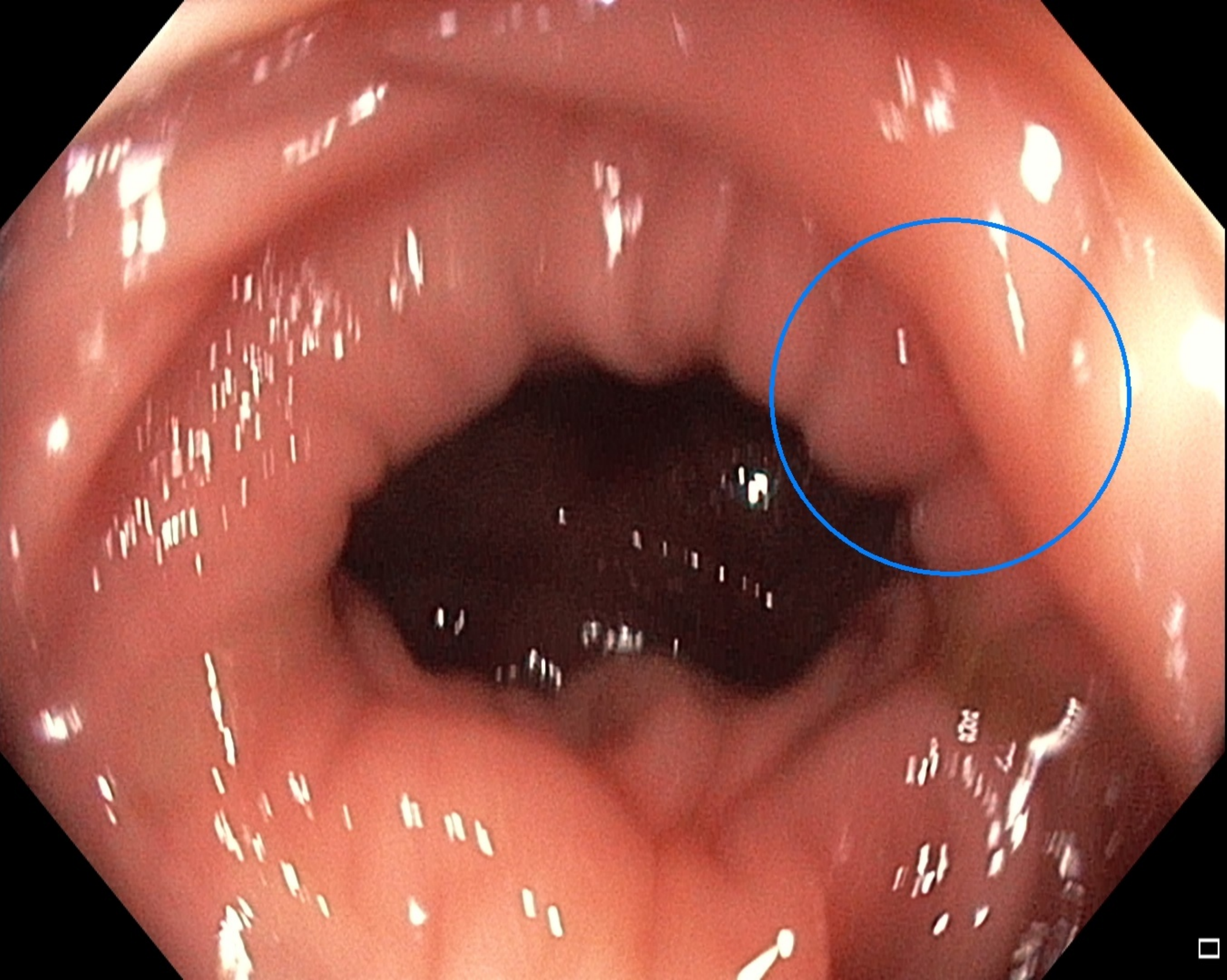}
    \caption{}
        \label{sfig:7}
  \end{subfigure}
  \hfill
  \begin{subfigure}[b]{0.24\textwidth}
    \includegraphics[width=\textwidth]{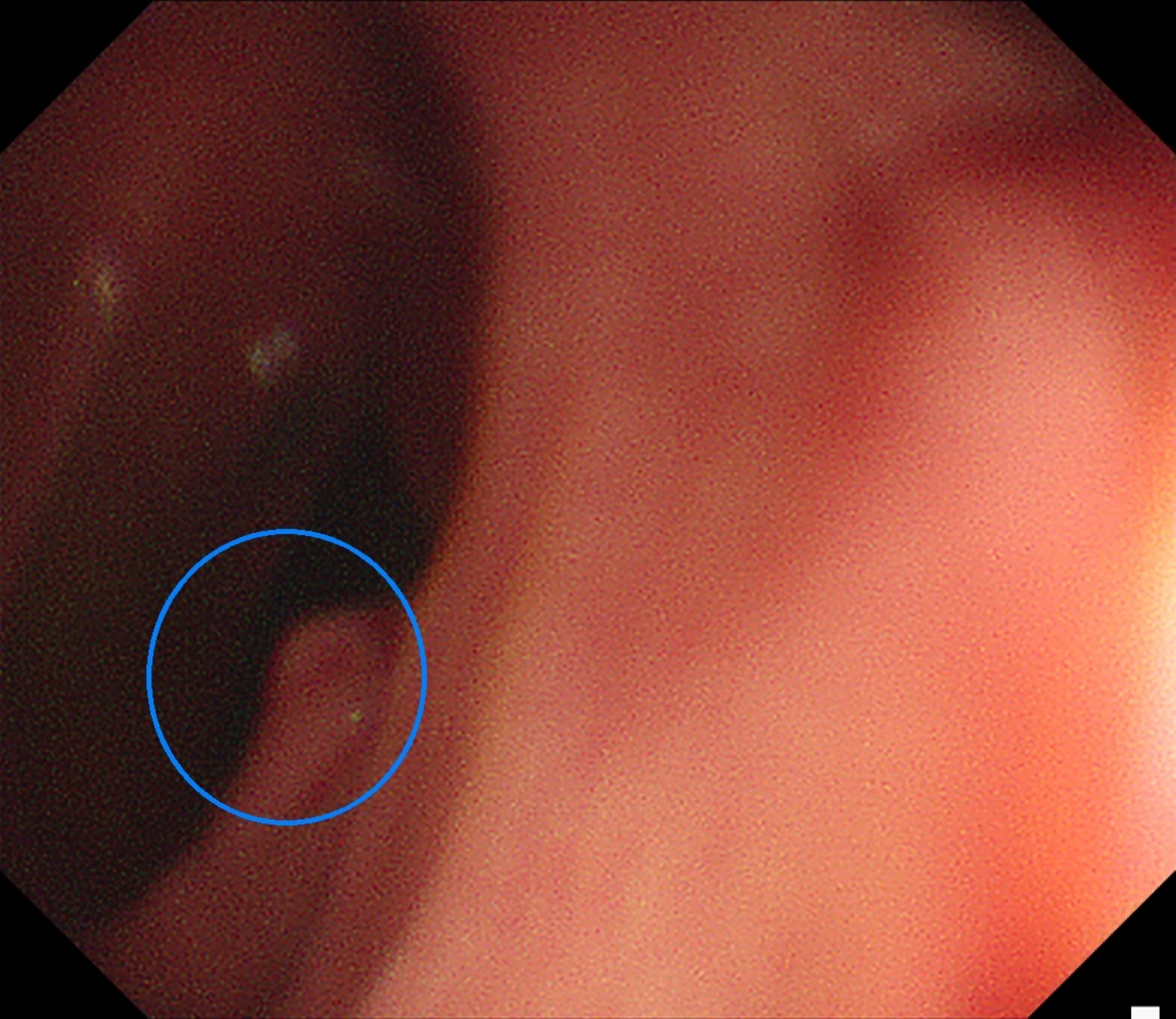}
    \caption{}
    \label{sfig:8}
  \end{subfigure}
  \caption{Sample images from the testing dataset, with results from the best performing model. White boxes are the ground truth annotations, blue ellipses are the model predictions. In the first row, examples of false negative polyps are shown: (\subref{sfig:1}) a small and distant polyp,  (\subref{sfig:2}) a polyp partially covered by water/bubbles, (\subref{sfig:3}) a polyp framed in blue light, (\subref{sfig:4}) a large polyp near the image boundary and overexposed. In the second row, examples of false positive detections are shown: (\subref{sfig:5}) the model activates on a artifact due to stain and motion blur, (\subref{sfig:6}) the model activates on a solid residue, (\subref{sfig:7}) the model activates on an area of the colonic mucosa that is not well inflated, (\subref{sfig:8}) the model activates on a dark and distant area of the colonic mucosa whose shape is similar to a polyp.}
  \label{fig:worsebest}
\end{figure}

Finally, in Figure \ref{fig:worsebest}, we display examples of false negatives and false positives from the test set, generated by the best performing model. To visually assess the performance on a whole video, we have uploaded a 60-minute colonoscopy video featuring 6 polyps, the longest in our test set, at \href{ https://figshare.com/s/fbb0834a21082984336c}{https://figshare.com/s/fbb0834a21082984336c} (with predictions marked in cyan and ground truth boxes in white). The image examples illustrate how the model struggles with small, occluded, or poorly imaged polyps, and generates false positives in areas that visually resemble polyps, often due to motion or suboptimal imaging. These observations persist throughout the entire video analysis, highlighting the importance of minimizing false positives throughout the entire procedure while maintaining high polyp recall.

\section*{Code availability}
To facilitate the process of downloading and exploring the dataset, we have made available a set of useful Python codes on our GitHub repository at \href{https://github.com/cosmoimd/real-colon-dataset}{https://github.com/cosmoimd/real-colon-dataset}. These scripts facilitate easy access to the data and assist in its analysis, enabling users to reproduce all the plots presented in this paper. Code for the training and testing of the polyp detection models can be found separately at \href{https://github.com/cosmoimd/DeepLearningExamples}{https://github.com/cosmoimd/DeepLearningExamples}.

\section*{Acknowledgements}
We wish to express our gratitude to Antonella Melina and Michela Ruperti for their assistance in coordinating the data acquisition and labeling process across the four data cohorts, to the Data Annotation Team at Cosmo Intelligent Medical Devices for their diligent efforts in annotating the dataset, to Erica Vagnoni and Stefano Bianchi for the development of the annotation software tool, and to Gabriel Marchese Aizenman for testing the GitHub repositories released with this work.

\section*{Author contributions statement}
C.B., P.S., and A.C. conceived the dataset, C.B. and P.S. curated the dataset and conducted the experiments. G.A., S.B., C.H., D.H, M.I., and A.M. collected the data. All authors reviewed the manuscript. 

\section*{Competing interests}
C.B., P.S., and A.C. are inventors of patents related to the subject of AI and are employees of a company manufacturing AI devices. C.H. is consultant for Medtronic and Fujifilm.

\bibliography{bibliography}

\begin{thebibliography}{10}
\urlstyle{rm}
\expandafter\ifx\csname url\endcsname\relax
  \def\url#1{\texttt{#1}}\fi
\expandafter\ifx\csname urlprefix\endcsname\relax\def\urlprefix{URL }\fi
\expandafter\ifx\csname doiprefix\endcsname\relax\def\doiprefix{DOI: }\fi
\providecommand{\bibinfo}[2]{#2}
\providecommand{\eprint}[2][]{\url{#2}}

\bibitem{sung2021global}
\bibinfo{author}{Sung, H.} \emph{et~al.}
\newblock \bibinfo{journal}{\bibinfo{title}{Global cancer statistics 2020:
  Globocan estimates of incidence and mortality worldwide for 36 cancers in 185
  countries}}.
\newblock {\emph{\JournalTitle{CA Cancer J. Clin.}}}
  \textbf{\bibinfo{volume}{71}}, \bibinfo{pages}{209--249}
  (\bibinfo{year}{2021}).

\bibitem{morgan2023global}
\bibinfo{author}{Morgan, E.} \emph{et~al.}
\newblock \bibinfo{journal}{\bibinfo{title}{Global burden of colorectal cancer
  in 2020 and 2040: incidence and mortality estimates from globocan}}.
\newblock {\emph{\JournalTitle{Gut}}} \textbf{\bibinfo{volume}{72}},
  \bibinfo{pages}{338--344} (\bibinfo{year}{2023}).

\bibitem{bretthauer2022effect}
\bibinfo{author}{Bretthauer, M.} \emph{et~al.}
\newblock \bibinfo{journal}{\bibinfo{title}{Effect of colonoscopy screening on
  risks of colorectal cancer and related death}}.
\newblock {\emph{\JournalTitle{N. Engl. J. Med.}}}
  \textbf{\bibinfo{volume}{387}}, \bibinfo{pages}{1547--1556}
  (\bibinfo{year}{2022}).

\bibitem{zorzi2023adenoma}
\bibinfo{author}{Zorzi, M. e.~a.}
\newblock \bibinfo{journal}{\bibinfo{title}{Adenoma detection rate and
  colorectal cancer risk in fecal immunochemical test screening programs: An
  observational cohort study}}.
\newblock {\emph{\JournalTitle{Ann. Intern. Med.}}}
  \textbf{\bibinfo{volume}{176}}, \bibinfo{pages}{303--310}
  (\bibinfo{year}{2023}).

\bibitem{dekker_advances_2018}
\bibinfo{author}{Dekker, E.} \& \bibinfo{author}{Rex, D.~K.}
\newblock \bibinfo{journal}{\bibinfo{title}{Advances in crc prevention:
  Screening and surveillance}}.
\newblock {\emph{\JournalTitle{Gastroenterology}}}
  \textbf{\bibinfo{volume}{154}}, \bibinfo{pages}{1970--1984}
  (\bibinfo{year}{2018}).

\bibitem{kaminski_optimizing_2020}
\bibinfo{author}{Kaminski, M.~F.}, \bibinfo{author}{Robertson, D.~J.},
  \bibinfo{author}{Senore, C.} \& \bibinfo{author}{Rex, D.~K.}
\newblock \bibinfo{journal}{\bibinfo{title}{Optimizing the quality of
  colorectal cancer screening worldwide}}.
\newblock {\emph{\JournalTitle{Gastroenterology}}}
  \textbf{\bibinfo{volume}{158}}, \bibinfo{pages}{404--417}
  (\bibinfo{year}{2020}).

\bibitem{cherubini_gorilla_2023}
\bibinfo{author}{Cherubini, A.} \& \bibinfo{author}{East, J.~E.}
\newblock \bibinfo{journal}{\bibinfo{title}{Gorilla in the room: Even experts
  can miss polyps at colonoscopy and how ai helps complex visual perception
  tasks}}.
\newblock {\emph{\JournalTitle{Dig. Liver Dis.}}}
  \textbf{\bibinfo{volume}{55}}, \bibinfo{pages}{151--153}
  (\bibinfo{year}{2023}).

\bibitem{ahmad2019artificial}
\bibinfo{author}{Ahmad, O.~F.} \emph{et~al.}
\newblock \bibinfo{journal}{\bibinfo{title}{Artificial intelligence and
  computer-aided diagnosis in colonoscopy: current evidence and future
  directions}}.
\newblock {\emph{\JournalTitle{Lancet Gastroenterol. Hepatol.}}}
  \textbf{\bibinfo{volume}{4}}, \bibinfo{pages}{71--80} (\bibinfo{year}{2019}).

\bibitem{berzin_position_2020}
\bibinfo{author}{Berzin, T. M. e.~a.}
\newblock \bibinfo{journal}{\bibinfo{title}{Position statement on priorities
  for artificial intelligence in gi endoscopy: a report by the asge task
  force}}.
\newblock {\emph{\JournalTitle{Gastrointest. Endosc.}}}
  \textbf{\bibinfo{volume}{92}}, \bibinfo{pages}{951--959}
  (\bibinfo{year}{2020}).

\bibitem{repici_efficacy_2020}
\bibinfo{author}{Repici, A. e.~a.}
\newblock \bibinfo{journal}{\bibinfo{title}{Efficacy of real-time
  computer-aided detection of colorectal neoplasia in a randomized trial}}.
\newblock {\emph{\JournalTitle{Gastroenterology}}}
  \textbf{\bibinfo{volume}{159}}, \bibinfo{pages}{512--520.e7}
  (\bibinfo{year}{2020}).

\bibitem{wallace_impact_2022}
\bibinfo{author}{Wallace, M. B. e.~a.}
\newblock \bibinfo{journal}{\bibinfo{title}{Impact of artificial intelligence
  on miss rate of colorectal neoplasia}}.
\newblock {\emph{\JournalTitle{Gastroenterology}}}  (\bibinfo{year}{2022}).

\bibitem{spadaccini_computer-aided_2021}
\bibinfo{author}{Spadaccini, M. e.~a.}
\newblock \bibinfo{journal}{\bibinfo{title}{Computer-aided detection versus
  advanced imaging for detection of colorectal neoplasia: a systematic review
  and network meta-analysis}}.
\newblock {\emph{\JournalTitle{Lancet Gastroenterol. Hepatol.}}}
  \textbf{\bibinfo{volume}{6}}, \bibinfo{pages}{793--802}
  (\bibinfo{year}{2021}).

\bibitem{biffi_novel_2022}
\bibinfo{author}{Biffi, C. e.~a.}
\newblock \bibinfo{journal}{\bibinfo{title}{A novel ai device for real-time
  optical characterization of colorectal polyps}}.
\newblock {\emph{\JournalTitle{NPJ Digit. Med.}}} \textbf{\bibinfo{volume}{5}},
  \bibinfo{pages}{84} (\bibinfo{year}{2022}).

\bibitem{bernal2012towards}
\bibinfo{author}{Bernal, J.}, \bibinfo{author}{Sánchez, J.} \&
  \bibinfo{author}{Vilariño, F.}
\newblock \bibinfo{journal}{\bibinfo{title}{Towards automatic polyp detection
  with a polyp appearance model}}.
\newblock {\emph{\JournalTitle{Pattern Recognit.}}}
  \textbf{\bibinfo{volume}{45}}, \bibinfo{pages}{3166--3182}
  (\bibinfo{year}{2012}).

\bibitem{silva2014toward}
\bibinfo{author}{Silva, J.}, \bibinfo{author}{Histace, A.},
  \bibinfo{author}{Romain, O.}, \bibinfo{author}{Dray, X.} \&
  \bibinfo{author}{Granado, B.}
\newblock \bibinfo{journal}{\bibinfo{title}{Toward embedded detection of polyps
  in wce images for early diagnosis of colorectal cancer}}.
\newblock {\emph{\JournalTitle{Int. J. Comput. Assist. Radiol. Surg.}}}
  \textbf{\bibinfo{volume}{9}}, \bibinfo{pages}{283--293}
  (\bibinfo{year}{2014}).

\bibitem{bernal2015wm}
\bibinfo{author}{Bernal, J. e.~a.}
\newblock \bibinfo{journal}{\bibinfo{title}{Wm-dova maps for accurate polyp
  highlighting in colonoscopy: Validation vs. saliency maps from physicians}}.
\newblock {\emph{\JournalTitle{Comput. Med. Imaging Graph.}}}
  \textbf{\bibinfo{volume}{43}}, \bibinfo{pages}{99--111}
  (\bibinfo{year}{2015}).

\bibitem{tajbakhsh2015automated}
\bibinfo{author}{Tajbakhsh, N.}, \bibinfo{author}{Gurudu, S.~R.} \&
  \bibinfo{author}{Liang, J.}
\newblock \bibinfo{journal}{\bibinfo{title}{Automated polyp detection in
  colonoscopy videos using shape and context information}}.
\newblock {\emph{\JournalTitle{IEEE Trans. Med. Imaging}}}
  \textbf{\bibinfo{volume}{35}}, \bibinfo{pages}{630--644}
  (\bibinfo{year}{2015}).

\bibitem{angermann2017towards}
\bibinfo{author}{Angermann, Q.} \emph{et~al.}
\newblock \bibinfo{title}{Towards real-time polyp detection in colonoscopy
  videos: Adapting still frame-based methodologies for video sequences
  analysis}.
\newblock In \emph{\bibinfo{booktitle}{Proc. 4th Int. Workshop CARE and 6th
  Int. Workshop CLIP, MICCAI 2017}}, \bibinfo{pages}{29--41}
  (\bibinfo{organization}{Springer}, \bibinfo{year}{2017}).

\bibitem{mesejo2016computer}
\bibinfo{author}{Mesejo, P.} \emph{et~al.}
\newblock \bibinfo{journal}{\bibinfo{title}{Computer-aided classification of
  gastrointestinal lesions in regular colonoscopy}}.
\newblock {\emph{\JournalTitle{IEEE Trans. Med. Imaging}}}
  \textbf{\bibinfo{volume}{35}}, \bibinfo{pages}{2051--2063}
  (\bibinfo{year}{2016}).

\bibitem{jha2020kvasir}
\bibinfo{author}{Jha, D.} \emph{et~al.}
\newblock \bibinfo{title}{Kvasir-seg: A segmented polyp dataset}.
\newblock In \emph{\bibinfo{booktitle}{Proc. 26th Int. Conf. MultiMedia
  Modeling, MMM 2020}}, \bibinfo{pages}{451--462} (\bibinfo{year}{2020}).

\bibitem{sanchez2020piccolo}
\bibinfo{author}{Sánchez-Peralta, L.~F.} \emph{et~al.}
\newblock \bibinfo{journal}{\bibinfo{title}{Piccolo white-light and narrow-band
  imaging colonoscopic dataset: a performance comparative of models and
  datasets}}.
\newblock {\emph{\JournalTitle{Appl. Sci.}}} \textbf{\bibinfo{volume}{10}},
  \bibinfo{pages}{8501} (\bibinfo{year}{2020}).

\bibitem{li2021colonoscopy}
\bibinfo{author}{Li, K.} \emph{et~al.}
\newblock \bibinfo{journal}{\bibinfo{title}{Colonoscopy polyp detection and
  classification: Dataset creation and comparative evaluations}}.
\newblock {\emph{\JournalTitle{PLoS ONE}}} \textbf{\bibinfo{volume}{16}},
  \bibinfo{pages}{e0255809} (\bibinfo{year}{2021}).

\bibitem{misawa2021development}
\bibinfo{author}{Misawa, M.} \emph{et~al.}
\newblock \bibinfo{journal}{\bibinfo{title}{Development of a computer-aided
  detection system for colonoscopy and a publicly accessible large colonoscopy
  video database (with video)}}.
\newblock {\emph{\JournalTitle{Gastrointest. Endosc.}}}
  \textbf{\bibinfo{volume}{93}}, \bibinfo{pages}{960--967}
  (\bibinfo{year}{2021}).

\bibitem{ma2021ldpolypvideo}
\bibinfo{author}{Ma, Y.}, \bibinfo{author}{Chen, X.}, \bibinfo{author}{Cheng,
  K.}, \bibinfo{author}{Li, Y.} \& \bibinfo{author}{Sun, B.}
\newblock \bibinfo{title}{Ldpolypvideo benchmark: a large-scale colonoscopy
  video dataset of diverse polyps}.
\newblock In \emph{\bibinfo{booktitle}{Proc. 24th Int. Conf. Med. Image Comput.
  Comput. Assist. Intervent., MICCAI 2021}}, \bibinfo{pages}{387--396}
  (\bibinfo{year}{2021}).

\bibitem{ali2023multi}
\bibinfo{author}{Ali, S.} \emph{et~al.}
\newblock \bibinfo{journal}{\bibinfo{title}{A multi-centre polyp detection and
  segmentation dataset for generalisability assessment}}.
\newblock {\emph{\JournalTitle{Sci. Data}}} \textbf{\bibinfo{volume}{10}},
  \bibinfo{pages}{75} (\bibinfo{year}{2023}).

\bibitem{nogueira-rodriguez_negative_2023}
\bibinfo{author}{Nogueira-Rodríguez, A.}, \bibinfo{author}{Glez-Peña, D.},
  \bibinfo{author}{Reboiro-Jato, M.} \& \bibinfo{author}{López-Fernández, H.}
\newblock \bibinfo{journal}{\bibinfo{title}{Negative samples for improving
  object detection—a case study in ai-assisted colonoscopy for polyp
  detection}}.
\newblock {\emph{\JournalTitle{Diagnostics}}} \textbf{\bibinfo{volume}{13}},
  \bibinfo{pages}{966} (\bibinfo{year}{2023}).

\bibitem{reverberi_experimental_2022}
\bibinfo{author}{Reverberi, C.} \emph{et~al.}
\newblock \bibinfo{journal}{\bibinfo{title}{Experimental evidence of effective
  human-ai collaboration in medical decision-making}}.
\newblock {\emph{\JournalTitle{Sci. Rep.}}} \textbf{\bibinfo{volume}{12}},
  \bibinfo{pages}{14952} (\bibinfo{year}{2022}).

\bibitem{ali2024assessing}
\bibinfo{author}{Ali, S.} \emph{et~al.}
\newblock \bibinfo{journal}{\bibinfo{title}{Assessing generalisability of deep
  learning-based polyp detection and segmentation methods through a computer
  vision challenge}}.
\newblock {\emph{\JournalTitle{Sci. Rep.}}} \textbf{\bibinfo{volume}{14}},
  \bibinfo{pages}{2032} (\bibinfo{year}{2024}).

\bibitem{bernal2017comparative}
\bibinfo{author}{Bernal, J.} \emph{et~al.}
\newblock \bibinfo{journal}{\bibinfo{title}{Comparative validation of polyp
  detection methods in video colonoscopy: results from the miccai 2015
  endoscopic vision challenge}}.
\newblock {\emph{\JournalTitle{IEEE Trans. Med. Imaging}}}
  \textbf{\bibinfo{volume}{36}}, \bibinfo{pages}{1231--1249}
  (\bibinfo{year}{2017}).

\bibitem{jha2020medico}
\bibinfo{author}{Jha, D.} \emph{et~al.}
\newblock \bibinfo{journal}{\bibinfo{title}{Medico multimedia task at mediaeval
  2020: Automatic polyp segmentation}}.
\newblock {\emph{\JournalTitle{arXiv preprint arXiv:2012.15244}}}
  (\bibinfo{year}{2020}).

\bibitem{hicks2021medico}
\bibinfo{author}{Hicks, S.} \emph{et~al.}
\newblock \bibinfo{title}{Medico multimedia task at mediaeval 2021:
  Transparency in medical image segmentation}.
\newblock In \emph{\bibinfo{booktitle}{Proceedings of MediaEval 2021 CEUR
  Workshop}} (\bibinfo{year}{2021}).

\bibitem{hicks2021medai}
\bibinfo{author}{Hicks, S.} \emph{et~al.}
\newblock \bibinfo{journal}{\bibinfo{title}{Medai: Transparency in medical
  image segmentation}}.
\newblock {\emph{\JournalTitle{Nordic Machine Intelligence}}}
  \textbf{\bibinfo{volume}{1}}, \bibinfo{pages}{1--4} (\bibinfo{year}{2021}).

\bibitem{hassan_artificial_2022}
\bibinfo{author}{Hassan, C.}, \bibinfo{author}{Balsamo, G.},
  \bibinfo{author}{Lorenzetti, R.}, \bibinfo{author}{Zullo, A.} \&
  \bibinfo{author}{Antonelli, G.}
\newblock \bibinfo{journal}{\bibinfo{title}{Artificial intelligence allows
  leaving-in-situ colorectal polyps}}.
\newblock {\emph{\JournalTitle{Clin. Gastroenterol. Hepatol.}}}
  \textbf{\bibinfo{volume}{20}}, \bibinfo{pages}{2505--2513.e4}
  (\bibinfo{year}{2022}).

\bibitem{participants_in_the_paris_workshop_paris_2003}
\bibinfo{author}{{Participants in the Paris Workshop}}.
\newblock \bibinfo{journal}{\bibinfo{title}{The {Paris} endoscopic
  classification of superficial neoplastic lesions: esophagus, stomach, and
  colon}}.
\newblock {\emph{\JournalTitle{Gastrointestinal Endoscopy}}}
  \textbf{\bibinfo{volume}{58}}, \bibinfo{pages}{S3--S43}
  (\bibinfo{year}{2003}).

\bibitem{schlemper_vienna_2000}
\bibinfo{author}{Schlemper, R.~J.} \emph{et~al.}
\newblock \bibinfo{journal}{\bibinfo{title}{The {Vienna} classification of
  gastrointestinal epithelial neoplasia}}.
\newblock {\emph{\JournalTitle{Gut}}} \textbf{\bibinfo{volume}{47}},
  \bibinfo{pages}{251--255} (\bibinfo{year}{2000}).

\bibitem{biffi_2023_real}
\bibinfo{author}{Biffi, C.} \emph{et~al.}
\newblock \bibinfo{journal}{\bibinfo{title}{Real-colon dataset}}.
\newblock {\emph{\JournalTitle{Figshare+}}}
  \url{https://doi.org/10.25452/figshare.plus.22202866.v2}
  (\bibinfo{year}{2024}).

\bibitem{lin2014microsoft}
\bibinfo{author}{Lin, T.-Y.} \emph{et~al.}
\newblock \bibinfo{title}{Microsoft coco: Common objects in context}.
\newblock In \emph{\bibinfo{booktitle}{Proc. 13th Eur. Conf. Comput. Vision,
  ECCV 2014}}, \bibinfo{pages}{740--755} (\bibinfo{year}{2014}).

\bibitem{liu2016ssd}
\bibinfo{author}{Liu, W.} \emph{et~al.}
\newblock \bibinfo{title}{Ssd: Single shot multibox detector}.
\newblock In \emph{\bibinfo{booktitle}{Proc. 14th Eur. Conf. Comput. Vision,
  ECCV 2016}}, \bibinfo{pages}{21--37} (\bibinfo{year}{2016}).

\end{thebibliography}

\end{document}